\documentclass{article}
\usepackage{amssymb}

\usepackage{graphicx}
\usepackage{amsmath}

\newtheorem{theorem}{Theorem}

\newtheorem{corollary}[theorem]{Corollary}

\newtheorem{definition}[theorem]{Definition}

\newtheorem{lemma}[theorem]{Lemma}

\newtheorem{proposition}[theorem]{Proposition}
\newtheorem{remark}[theorem]{Remark}

\newenvironment{proof}[1][Proof]{\textbf{#1.} }{\ \rule{0.5em}{0.5em}}
\newenvironment{pf}[1][Proof of Proposition]{\textbf{#1} }{\ \rule{0.5em}{0.5em}}

\newcommand {\dt}[1]{\overset{.}{#1}}
\newcommand{\pr}[1]{\frac{\partial}{\partial #1}}

\newcommand{\Imm}{\mbox{{\rm Im}}}
\newcommand{\Tor}{\mbox{{\rm Tor}}}

\newcommand{\Sym}{\mbox{{\rm Sym}}}
\newcommand {\Ker}{\mbox{{\rm Ker}}}

\newcommand{\dbar}{\bar \partial}

\begin{document}
\title{Yang-Mills theories in dimensions 3,4,6,10 and Bar-duality} 
\author{M.Movshev\\IHES\\Bures-sur-Yvette\\France}
\maketitle
\begin{abstract}
In this note we give a homological explanation of ``pure spinors'' in YM theories with minimal amount of supersymmetries.
We construct A$_{\infty}$ algebras $A$ for every dimension $ D=3,4,6,10$, which for $D=10$ coincides with homogeneous coordinate ring of pure spinors with coordinate $\lambda^{\alpha}$. These algebras are Bar-dual to Lie algebras generated by supersymmetries, written in components.
The algebras  have a finite number of higher multiplications. It is   typical for generic A$_{\infty}$ algebra. The main result of the present note is that in dimension $D=3,6,10$  the algebra $A\otimes \Lambda[\theta^{\alpha}]\otimes Mat_n$ with a  differential  $d$ is equivalent to Batalin-Vilkovisky algebra of minimally supersymmetric YM theory in dimension $D$ reduced to a point. This statement can be extended to nonreduced theories.
\end{abstract}

\section{Introduction}
In \cite{Howe} P.Howe and in  \cite{Berk1} N.Berkovits suggested to formulate $N=1,D=10$ YM theory in terms of pure spinors. Suppose $\Gamma$ is 10-dimensional gamma matrices. $\lambda^1,\dots,\lambda^{16}$ are coordinates on $16$-dimensional spinor representation of $\mathfrak{so}(10)$. Introduce an algebra of pure spinors:
\begin{equation}\label{E:sgdbc}
A=\mathbb{C}[\lambda^1,\dots,\lambda^{16}]/(\Gamma^i_{\alpha \beta}\lambda^{\alpha}\lambda^{\beta})
\end{equation} This algebra can be regarded as a coordinate ring of a homogeneous manifold $SO(10,\mathbb{R})/U(5)$. It also has a description as of a connected component of isotropic 5-dimensional flags $V\subset \mathbb{C}^{10}$ with respect to $\mathbb{C}$-linear (not hermitian) symmetric  $\mathfrak{so}(10)$-invariant bilinear form $(.,.)$.

Define $B$ as the algebra
\begin{equation}
B=A\otimes \Lambda[\theta^{1},\dots,\theta^{16}]\otimes C^{\infty}(\mathbb{R}^{10})
\end{equation}
where $C^{\infty}(\mathbb{R}^{10})$ are smooth functions on $\mathbb{R}^{10}$. The algebra $B$ has a remarkable differential
\begin{equation}\label{D:hsuwe}
d=\lambda^{\alpha}\pr{\theta^{\alpha}}+\Gamma^{i}_{\alpha \beta}\lambda^{\alpha}\theta^{\alpha}\pr{x^i}
\end{equation}

The operator satisfies $d^2=0$. The generators of the algebra of supersymmetry act on $B$ by the vector fields
\begin{equation}
\pr{\theta^{\alpha}}-\Gamma^{i}_{\alpha \beta}\theta^{\alpha}\pr{x^i}
\end{equation}
\begin{definition}
A Maurer-Cartan equation for differential graded Lie algebra $(\mathfrak{g},d)$ is
\begin{equation}\label{E:hsfs}
da+\frac{1}{2}\{a,a\}=0, \quad a\in \mathfrak{g}
\end{equation}
\end{definition}
The algebra $B\otimes Mat_n$ is also a differential algebra.
\begin{proposition}
There is a one-to -one correspondence between complexified space of gauge-equivalent classes of $N=1$, $D=10$ YM equation with a gauge group $U(n)$ and classes of solutions of Maurer-Cartan equation \ref{E:hsfs} for the algebra $B\otimes Mat_n$.
\end{proposition}

Berkovits introduced a linear functional $tr:B\otimes Mat_n\rightarrow \mathbb{C}$ is which defined for sufficiently fast decaying at infinity elements of $B\otimes Mat_n$. The functional $tr$ is $d$-closed: $tr(da)=0$ for all $a\in B\otimes Mat_n$. Consider elements of $B\otimes Mat_n$ as fields of some field theory . Define a Lagrangian by the formula:
\begin{equation}
S(A)=tr(\frac{1}{2}ada+\frac{1}{6}a^3)
\end{equation}
Conjecturally this field theory is equivalent to $N=1,D=10$ YM theory.

In this note we will try generalize  the approach of N.Berkovits to YM theory with smaller amount of supersymmetries. At a first glance one can see no room in his construction for such generalization. A careful analysis of supersymmetry algebra enables us to do this . The main observation that the coordinate ring $A$ of pure spinor is Koszul dual to supersymmetry algebra of D=10 N=1 YM theory written in components was made in \cite{MSch}. This statement could be promoted to a definition  of algebra pure spinors for YM theories with lesser supersymmetries. 
In such setup we use more general Bar-duality as a replacement of Koszul duality.

Few words about Bar-duality are in order.
Suppose $\mathfrak{g}$ is a positively graded Lie algebra. Its Lie algebra cohomology with trivial coefficients $H(\mathfrak{g},\mathbb{C})$ is cohomology of exterior algebra $\Lambda[\mathfrak{g}^*]$, equipped with differential $d$. On linear generators it is equal $(dc)(g_1,g_2)=c([g_1,g_2])$. The definition of cohomology can be extended to differential graded Lie algebras and even L$_{\infty}$-algebras. By construction the cohomology $H(\mathfrak{g},\mathbb{C})$ is graded commutative algebra. In fact it carries higher multiplications, making it C$_{\infty}$-algebra.

Suppose $A=\bigoplus_{i\geq0}A_i$ is a graded commutative algebra. We assume that $A_0=\mathbb{C}$, thus there is a homomorphism $\epsilon:A\rightarrow A_0$. Define Hochschild cohomology of $A$ with trivial coefficients $H(A,\mathbb{C})$ as cohomology of complex $C^n=Hom(A^{\otimes n},\mathbb{C})$. The differential is defined by the rule:
\begin{equation}
\begin{split}
&(dc)(a_0,\dots,a_n)=\epsilon(a_0)c(a_1,\dots,a_n)+\\
&+\sum_{i=0}^{n-1}(-1)^{i-1}c(a_0,\dots,a_ia_{i+1},\dots,a_n)+\\
&(-1)^{n+1}c(a_0,\dots,a_{n-1})\epsilon(a_n)
\end{split}
\end{equation}
It turns out that commutativity of $A$ ensures that $H(A,\mathbb{C})$ is in fact a universal enveloping algebra of some Lie algebra $HQ(A,\mathbb{C})$. The Lie algebra $HQ(A,\mathbb{C})$ carry a higher multiplications, which make it L$_{\infty}$-algebra. One can extend Hochschild cohomology to differential commutative algebras, or even to C$_{\infty}$-algebras. The main fact which we will use extensively throughout present paper can be formulated as a theorem:
\begin{proposition}(Bar-duality)
For any positively graded Lie algebra $\mathfrak{g}$ we have $HQ(H(\mathfrak{g},\mathbb{C}),\mathbb{C})=\mathfrak{g}$. For any positively graded commutative algebra $A$ we have $H(HQ(A,\mathbb{C}),\mathbb{C})=A$. In the above isomorphisms $HQ(A,\mathbb{C})$ is considered as L$_{\infty}$-algebra, $H(\mathfrak{g},\mathbb{C})$ as C$_{\infty}$-algebra.
\end{proposition}

The above classical theorem is valid in a broader generality, when $\mathfrak{g}$ is L$_{\infty}$-algebra and $A$ is C$_{\infty}$-algebra (see \cite{MKos} for details).
The  theorem asserts that no information is lost upon transition from algebra $\mathfrak{g}$ to $H(\mathfrak{g},\mathbb{C})$ if higher multiplications are taken into account. In particular if $\mathfrak{g}=L$ is an algebra generated by  supersymmetries for theory written in components, then such algebra carries much information about YM theory (if not all). As a consequence the mentioned theorem the  Bar-dual algebra $H(L,\mathbb{C})$ carries the same amount of information. The algebra $H(L,\mathbb{C})$ is commutative, which could be used in our advantage because it makes a link to algebraic geometry.
 
In this note we made a computation of ``pure spinors'' for $N=1$ supersymmetric YM theories in dimensions $3,4,6,10$. The case $D=10$ was treated in \cite{MSch2}.

The algebras of ``pure spinors'' in smaller dimensions  have additional structure: All of them are A$_{\infty}$ algebras (see \cite{MSch2} for definition and discussion). More precisely they are C$_{\infty}$-algebras but A$_{\infty}$ condition will be sufficient for us. Shortly A$_{\infty}$ condition is a generalization of associativity condition for ordinary algebra. An A$_{\infty}$ algebra carries for any $n\geq 1$ a (possibly zero) $n$-ary operation $\mu_n:A^{\otimes n}\rightarrow A$. There is a quadratic equation relating the maps in this collection, which if $\mu_n=0, n\neq 2$ becomes associativity condition for $\mu_2$.

In our case almost all $\mu_n$ are equal to zero.

{\bf 1} $D=10,N=1$ , only $\mu_2$ survives: this is a multiplication in the algebra of pure spinors \ref{E:sgdbc}.

{\bf 2} $D=6,N=1$, The algebra $A$ is a direct sum of coordinate ring $A^t=\bigoplus_{i \geq 0}H^0(\mathbf{P}^3\times \mathbf{P}^1,{\cal O}(i,i))$ and an ideal $I=\bigoplus_{i \geq 0}H^0(\mathbf{P}^3\times \mathbf{P}^1,{\cal O}(i,i+2))$, with trivial multiplication on the ideal. Besides ordinary multiplication there is operation $\mu_4$, which is a Hochschild 4-cocycle $\mu_4:A^{t\otimes 4}\rightarrow I$.

{\bf 3} $D=4$, $N=1$ , the underlying manifold for the algebra $A=H(L,\mathbb{C})$ is $\mathbf{P}^1\cup \mathbf{P}^1$. The algebra $A$ has operations $\mu_2,\mu_3,\mu_4,\mu_5$.

{\bf 4 } $D=3$, $N=1$ The algebra itself has no nice geometric description, but closely related to superspace formulation of YM theory. The algebra is finite dimensional. It has two operations $\mu_2,\mu_5$.

The A$_{\infty}$ algebras briefly sketched in item 1-4 shall be denoted by $A$. Denote $S$ the spinor representation in irreducible SUSY data (see Appendix) with  $dim(V)=D$. The space  has a basis   $\tilde\theta^1,\dots,\theta^{dim(S)}$. Denote by  $B$ the tensor product:

\begin{equation}
B=A\otimes \Lambda[S]\otimes C^{\infty}(V)
\end{equation}
The algebra $B$ is suppose to carry a deformation of differential \ref{D:hsuwe}. Thus we get a differential A$_{\infty}$ algebra.

Any A$_{\infty}$ algebra defines L$_{\infty}$ algebra by (skew)symmetrization of arguments in operations $\mu_n$. There is  generalization of Maurer-Cartan equation \ref{E:hsfs} for A$_{\infty}$ and  L$_{\infty}$ algebras, which for the later has a form
\begin{equation}\label{E:jhatrw}
\mu_1(a)+\frac{1}{2}\mu_{2}(a,a)+\dots+\frac{1}{n!}\mu_n(a,\dots,a)+\dots=0
\end{equation}
\begin{proposition}
The complexified set of gauge equivalence classes of solutions of $N=1$ , $dim=D=3,6,10$ YM equation with a gauge group $U(n)$ is isomorphic to classes of Maurer-Cartan equation \ref{E:jhatrw} for the algebra $B\otimes Mat_n$
\end{proposition}

There is a trace functional $tr:B(D)\otimes Mat_n\rightarrow \mathbb{C}$, which is $d$-closed. Together with $\mu_2$ it defines $d$-closed inner product $(a,b)=tr\mu_2(a,b)$. On the space of fields $B(D)\otimes Mat_n$ define a Lagrangian
\begin{equation}
S(a)=\frac{1}{2!}(a,\mu_1a)+\frac{1}{3!}(a,\mu_2(a,a))+\dots
\end{equation}

It is very likely that the theory defined by this Lagrangian is completely equivalent to $N=1$ $dim=D$ YM theory
To achieve a cleaner algebraic picture we reduce all the theories to a point.
We shall prove
\begin{proposition}
The algebra $A\otimes \Lambda[S]$ is quasiisomorphic to Bar-dual to BV algebra of D-dimensional N=1 supersymmetric YM theory reduced to a point.
\end{proposition}


In this note we use Einstein summation convention over repeated indecies.  Greek letter are used for spinor, Latin for vector indecies. We fix a diagonal metric tensor $\delta^{ij}$ and its inverse $\delta_{ij}$. In it presence of such metric  we make no distinction of upper and lower vector indecies. All linear spaces are considered over field of complex numbers.
\section{Algebra of supersymmetries}
\subsection{Notations and definitions}
In this note the key object is $YM$ algebra. It was defined in \cite{MSch},\cite{MSch2}. However the definition given in the  mentioned papers is a bit too general. To make it more suitable for our purposes we need a definition of spinors and SUSY data which is provided in the Appendix for the readers convenience. 
\begin{definition}
Let $V$ be $n$-dimensional linear space over complex numbers. It is  equipped with a symmetric nondegenerate dot-product. Let $S$ be a spinor representation of $Spin_n$ which is involved in the definition of a  SUSY data ( see Appendix). Suppose $A_1,\dots,A_n$ is an orthonormal basis of $V$, $\psi^{\alpha}$ $\alpha=1,\dots, dim(S)$ is a basis of $S$.  $YM$ algebra is a quotient of a free Lie algebra $Free<A_1,\dots,A_n,\psi^{1},\dots,\psi^{dim(S)}>$. The parity of $A_i$ is even and of $\psi^{\alpha}$ is odd. The ideal is generated by relations
\begin{align}
&\tilde A_m=[A_{i},[A_{i},A_{m}]]-\frac{1}{2}\Gamma_{\alpha \beta}^m\{\psi^{\alpha},\psi^{\beta}\}\label{E:jhda} \\
&\tilde \psi_{\alpha}=\Gamma_{\alpha \beta}^i[A_i,\psi^{\beta}]\label{E:jhda2}
\end{align}
\end{definition}
Denote $U(YM)$ the universal enveloping of $YM$.
\begin{definition}\label{E:hsdcfd}
It is convenient to introduce a grading on $YM$ and  $U(YM)$ such that $deg(A_i)=2$, $deg(\psi^{\alpha})=3$.
\end{definition}
\begin{definition}
In a (Lie) algebra $\mathfrak{g}$ the center is a set of all elements $x$ such that $[x,a]=0$ for all $a\in \mathfrak{g}$.
\end{definition}
\begin{proposition}\label{P:xccaed}
The center of $U(YM)$ is equal to $\mathbb{C}$, the center of $YM$ is equal to zero for $dim(V) + dim(S) \geq 3$.
\end{proposition}
\begin{proof}
The proof will be given in the forthcoming papers \cite{MSch5} \cite{MSch6}.
\end{proof}

We need to describe a convention of indexing cohomology of differential graded Lie algebra $\mathfrak{g}$, which is equipped with differential $d$ of degree $-1$. 
\begin{definition}
The exterior algebra $\Lambda[\mathfrak{g}^*]$ is bigraded: the auxiliary grading $adeg(\Lambda^i[\mathfrak{g}^*])$ is equal to $i$. The tensor powers of a graded space ($\mathfrak{g}^*$ in our case) have induced grading, which is the sum of gradings of tensor components. We call such a homogeneous grading $hdeg$(we use a convention that dualization does not change $hdeg$ of a homogeneous graded component). The topological grading is $deg=adeg+hdeg$. The differential $\partial$ on $\Lambda[\mathfrak{g}^*]$ is defined  by the Lie bracket $[.,.]$ and differential $d$ on $\mathfrak{g}$. It  satisfies $deg(\partial a)=deg(a)+1$. This will be the main grading in cohomology groups. If we assume that $d=0$ then the cohomology acquire an additional index- $hdeg$.
\end{definition}
Suppose $A$ is a Koszul algebra ( see \cite{MSch2} for definition), $M$ is $A$ bimodule. In the product $A_1\otimes A_1^*\subset A\otimes A^!$ we have an element $e=\sum_ie_i\otimes e^i$, where $e_i$ is a basis of $A_1$, $e^i$ is the dual basis. We have $e^2=0$ (see \cite{Qalg} for details). Define a cohomological grading on $A^!\otimes M$ as grading of $A^!$ tensor component. For $x\in M\otimes A^!$ define $dx=\{e,x\}$, where $\{.,.\}$ stands for graded commutator. 

Let $HH(A,M)$ be Hochschild cohomology group of $A$ with coefficients in $M$ (see \cite{MSch2} for definition and references).
\begin{proposition}\label{P:vdsyh}
The groups  $HH(A,M)$ are isomorphic to cohomology groups of $(M\otimes A^!,d)$
\end{proposition}
\begin{proof}
See \cite{MKos} for the proof.
\end{proof}

For future reference we formulate without the proof the following
\begin{proposition}\label{P:ghsfa}
Suppose $\mathfrak{g}$ is a positively graded Lie algebra generated by $\mathfrak{g}_1$. Then the L$_{\infty}$ algebra $H(\mathfrak{g},\mathbb{C})$ is generated by  $H^{2,1}(\mathfrak{g},\mathbb{C})$.
\end{proposition}
the proof will appear in \cite{MKos}.
\begin{definition}
Denote $A=H(\mathfrak{g},\mathbb{C})$. Denote $A^t$ subalgebra generated by $H^{2,1}(\mathfrak{g},\mathbb{C})$.
\end{definition}

\begin{proposition}\label{E:mkdgaf}
Suppose $L=\bigoplus_{i\geq1}L_i$ is a graded Lie algebra with Koszul universal enveloping $U(L)$. Denote $A$ a commutative algebra , which is Koszul dual to $U(L)$. Denote $\lambda^1,\dots,\lambda^{dim L_1}$ a generating set, $\theta^1,\dots,\theta^{dim L_1}$ a basis of $L_1^*$. Let $M=\bigoplus_{i\geq2}L_i$. Then the cohomology $H(M,\mathbb{C})$ is isomorphic to cohomology of the Koszul complex $(A\otimes \Lambda[L_1^*],\lambda^{\alpha}\pr{\theta^{\alpha}})$. One can say more: the A$_{\infty}$ algebra of $H(M,\mathbb{C})$ is quasiisomorphic to $(A\otimes \Lambda[L_1^*],d)$. Another way of phrasing this is to say that $(A\otimes \Lambda[L_1^*],d)$ is quasiisomorphic to Bar-dual of $U(M)$ The cohomology $H(M,\mathbb{C})$ carry a bigrading with respect to topological and homogeneous indecies. The algebra $A\otimes \Lambda[L_1^*]$ is also bigraded. The generating space $A_1\subset A$ has bidegree $(2,1)$. The generating space $L_1^*$ has bigrading $(1,1)$.
\end{proposition}
\begin{proof}
See \cite{MSch2}
\end{proof}
\begin{proposition}\label{P:lswcd}
Suppose $L$ is a positively graded Lie algebra with cohomology ring $H(L,\mathbb{C})$. If the auxiliary grading of the nonzero graded components of $H(L,\mathbb{C})$ is equal to $0,1$ then the algebra $L$ is free. The linear space of elements of auxiliary degree one is dual to the linear space of generators of  $L$.

More generally if the graded component of auxiliary degree two is not equal to zero then its dual  space is isomorphic to the minimal generating space of ideal of relations.
\end{proposition}
\begin{proof}
See \cite{McL}.
\end{proof}

%

Denote $Der(A)$ the  Lie algebra  of graded derivation of (Lie) algebra $A$. For any element $x\in A$ define a derivation $ad(x)$ by the formula
\begin{equation}
ad(x)a=[x,a]
\end{equation}
\begin{corollary}
As a corollary of proposition \ref{P:xccaed} we obtain that the  map $ad:YM\rightarrow Der(YM)$ is injective.
\end{corollary}

It is a matter of a simple check the following

\begin{proposition}
The image $\Imm\  ad$ is an ideal in $Der(A)$
\end{proposition}

\subsection{Cohomology of the algebra of supersymmetries}
\begin{definition}
A Fierz identity on $\Gamma$-matrices from SUSY data is
\begin{equation}\label{E:afdnc}
\Gamma_{\alpha\beta}^i\Gamma_{\gamma\delta}^i+\Gamma_{\alpha\gamma}^i\Gamma_{\beta\delta}^i+\Gamma_{\alpha\delta}^i\Gamma_{\beta\gamma}^i=0
\end{equation}
\end{definition}
This identity holds for low dimensional spinor representation and does not follows from defining identity \ref{E:javfx}.
\begin{proposition}\label{E:hdsvx}
If SUSY data satisfies Fierz identity \ref{E:afdnc} then it is irreducible. The dimension of $V$ could be equal to $3,4,6,10$
\end{proposition}
\begin{proof}
The proof will be given in section \ref{S:xdqde}
\end{proof}


\begin{proposition}
Suppose a SUSY data satisfies Fierz identity \ref{E:afdnc}. As usual it satisfies \ref{E:javfx}.
Then the algebra $YM$  has a remarkable set of differentiations $\theta_{\alpha}\in S^*$ called supersymmetries. They are defined by the formulas
\begin{equation}\label{R:kahdb}
\begin{split}
&\theta_{\alpha}(A_i)=\Gamma_{\alpha\beta i}\psi^{\beta}\\
&\theta_{\alpha}(\psi^{\beta})=-\frac{1}{2}\Gamma_{\alpha}^{\beta ij}[A_i,A_j]
\end{split}
\end{equation}

\end{proposition}
\begin{proof}
Given in section \ref{S:xdqde}
\end{proof}

In \cite{MSch2} we constructed a free resolution for the algebra $YM$. It is easy to extend the action of supersymmetry algebra to the resolution. Recall that this resolution is a free Lie algebra $R$ with
\begin{equation}
A_1,\dots,A_{dim(V)},A^{*1},\dots,A^{*dim(V)},\psi^{1},\dots,\psi^{dim(S)},\psi^*_{1},\dots,\psi^*_{dim(S)},c^*
\end{equation} as a generators. The parity of a generator $x^*$ from above list is opposed to the parity of $x$. The algebra $R$ is equipped with a differential

\begin{equation}\label{E:yunbr}
\begin{split}
&Q(A_i)=0\\
&Q(\psi^{\alpha})=0\\
&Q(c^*)=[A_i,A^{*i}]+[\psi^{\alpha},\psi_{\alpha}^{*}]\\
&Q(A^{*m})=-[A_{i},[A_{i},A_{m}]]+\frac{1}{2}\Gamma^m_{\alpha\beta}\{\psi^{\alpha},\psi^{\beta}\}=-\tilde A^m\\
&Q(\psi_{\alpha})=-\Gamma_{\alpha\beta}^i[A_i,\psi^{\beta}]=-\tilde \psi_{\alpha}\\
\end{split}
\end{equation}

\begin{proposition}
Define the action of $\theta_{\alpha}$ on the resolution $R$. $\theta_{\alpha}$ acts on $A_i, \psi^{\beta}$ by the formula \ref{R:kahdb}, on $\psi_{\beta}^*$ by \ref{E:safxg} and on $ A^*_i$ by \ref{E:csdsj} (with $\tilde{\ }$ replaced by $\ ^*$) and  on $c^*$ by the formula
\begin{equation}
\theta_{\alpha}c^*=0
\end{equation}
Then $\theta_{\alpha}$ commute with $Q$.
\end{proposition}
\begin{proof}
All we need to check is that $\theta_{\alpha}([A_i,A^{*i}]+[\psi^{\alpha},\psi^*_{\alpha}])=0$. This simple exercises on $\Gamma$-matrix manipulation is left to the reader.
\end{proof}
\begin{proposition}
Derivations $\theta_{\alpha}$ acting on the Lie algebra $L$ satisfy
\begin{equation}\label{P:xsjhfi}
\{\theta_{\alpha},\theta_{\beta}\}=-2\Gamma_{\alpha\beta}^iad(A_i)
\end{equation}
\end{proposition}
\begin{proof}
If the reader got this far, he should be comfortable enough  dealing with $\Gamma$ matrices and should manage the proof by himself. It contains no tricks. One should use the main  and the Fierz  identities.
\end{proof}

The elements $\theta_{\alpha}$ generate a subalgebra in $Der(YM)$ which we denote by $L$. It follows immediately from proposition \ref{P:xsjhfi} and relations \ref{R:kahdb} that $YM$ is an ideal in $L$ and $L/YM=S^*$ is an odd abelian Lie algebra. The grading introduced in definition \ref{E:hsdcfd} can be extended to $L$. We set $deg \theta_{\alpha}=1$. 

The algebra $L$ along with $YM$ will be  central  in the present note.

Our first task will be computation of cohomology $H^{\bullet}(L)$ of Lie algebra $L$ with trivial coefficients. The cohomology will be a commutative algebra, A connection of $YM$-theory  with algebraic geometry will be made by means of $H^{\bullet}(L)$.

We  need to remind the reader the results of computation of cohomology of $YM$ from \cite{MSch2}.
\begin{proposition}
The following isomorphisms hold:
\begin{equation}
\begin{split}
&H^{0,0}(YM,\mathbb{C})=H^{11,8}(YM,\mathbb{C})=\mathbb{C}\\
&H^{3,2}(YM,\mathbb{C})=V\quad H^{4,3}(YM,\mathbb{C})=S^*\\
&H^{7,5}(YM,\mathbb{C})=S\quad H^{8,6}(YM,\mathbb{C})=V
\end{split}
\end{equation}
There is also a nondegenerate pairing
\begin{equation}\label{E:svdh}
H^{i,j}(YM,\mathbb{C})\otimes H^{11-i,8-j}(YM,\mathbb{C})\rightarrow H^{11,8}(YM,\mathbb{C})
\end{equation}
induced by multiplication in cohomology.
\end{proposition}

Consider a complex
\begin{equation}\label{E:mxbdgdi}
\begin{split}
&N=\Sym^{i}(S^{*})\overset{\eta_0}{\leftarrow} \Sym^{i-2}(S^{*})\otimes V\overset{\eta_1}{\leftarrow} \Sym^{i-3}(S^{*})\otimes S\overset{\eta_2}{\leftarrow}\\
&  \overset{\eta_2}{\leftarrow} \Sym^{i-5}(S^{*})\otimes S^{*}\overset{\eta_3}{\leftarrow}  \Sym^{i-6}(S^{*})\otimes V\overset{\eta_4}{\leftarrow}  \Sym^{i-8}(S^{*})
\end{split}
\end{equation}
with $N_0(i)=\Sym^{i}(S^{*})$, $N_{1}(i)=\Sym^{i}(S^{*})\otimes V$, $N_{2}(i)=\Sym^{i}(S^{*})\otimes S$, $N_{3}(i)=\Sym^{i}(S^{*})\otimes S^*$,$N_{4}(i)=\Sym^{i}(S^{*})\otimes V$, $N_{5}(i)=\Sym^{i}(S^{*})$.

Introduce notations for free generators of $\Sym(S^*)$-modules $N_k$. We use brackets $<x_1,\dots,x_k>$ to denote a linear span of symbols $x_1,\dots,x_k$. Then 
\begin{equation}\label{E:qrshbdh}
\begin{split}
&N_{0}(0)=<1>\\
&N_{1}(0)=<A^*_1,\dots,A^*_{dim(V)}>\\
&N_{2}(0)=<\psi^*_1,\dots,\psi^*_{dim(S)}>\\
&N_{3}(0)=<\mu^{*1},\dots,\mu^{*dim(S)}>\\
&N_{1}(0)=<B^{*1},\dots,B^{*dim(V)}>\\
&N_{5}(0)=<\omega^*>
\end{split}
\end{equation} 
Denote linear generators of $\Sym(S^*)$ by $<\theta^{*1},\dots,\theta^{*dim(S)}>$
The maps $\eta_i$
\begin{align}
&\eta_0A^{*i}=-2\Gamma^i_{\alpha\beta}\theta^{*\alpha}\theta^{*\alpha}\\
&\eta_{1}\psi^{*}_\alpha=\Gamma_{i\alpha\beta}\theta^{*\beta}A^{*i}\\
&\eta_{2}\mu^{*\alpha}=\Gamma_i^{\alpha\beta}\Gamma^i_{\gamma\delta}\theta^{*\gamma}\theta^{*\delta}\psi^*_{\beta}+\theta^{*\alpha}\theta^{*\beta}\psi^*_{\beta}\label{E:rfhdww}\\
&\eta_{3}B^{*i}=\Gamma_{\alpha\beta}^i\theta^{*\alpha}\mu^{*\beta}\\
&\eta_4\omega^*=\Gamma_{\alpha\beta}^i\theta^{*\alpha}\theta^{*\beta}B^{*i}
\end{align}

\begin{proposition}\label{P:dxcj}
There is an isomorphism $H^{i,j}(L,\mathbb{C})=\bigoplus_{2j+k=i}H^k(N(j))$

\end{proposition}
\begin{proof}
The proof will be given in section \ref{S:xdqde}
\end{proof}

\section{D=10 N=1 YM}

This case is pretty well understood. Let $A=\Sym(S^*)/\Gamma^i_{\alpha\beta}\lambda^{\alpha}\lambda^{\beta}$. The cohomology of Berkovits(Koszul) complex $A\otimes \Lambda[\theta^1,\dots,\theta^{16}]$ with differential $\lambda^{\alpha}\pr{\theta^{\alpha}}$ compute the groups $\Tor_i^{\Sym(S^*)}(A,\mathbb{C})$(see \cite{MSch2}). The later groups were computed in \cite{Berk1} (more rigorously in \cite{MSch2}). These groups coincide with $N_i(0)$ from \ref{E:qrshbdh}. The group  $\Tor_i^{\Sym(S^*)}(A,\mathbb{C})$ is a space of  generators of $i$-th term  a $\Sym(S^*)$-free minimal graded resolution of $A$.  The grading and $Spin(10)$-equivariance condition completely fix the differentials $d_i$ in the resolution making them equal to $\eta_i$. We conclude that the complex $A$ is acyclic in all terms but zero, in which it cohomology is equal to  $A$.

The constructed resolution  was first time introduced in \cite{CR} in a different context.

In the table below we  illustrated the content of the cohomology of the complex $A$ as of representations of $Spin(10)$. The $j$-th row corresponds to cohomology of the complex $N(j)$.
 
We characterize a representation of $Spin(10)$ by coordinates of its highest weight. In particular $S^*=[0,0,0,0,1]$, $S=[0,0,0,1,0]$ and $V=[1,0,0,0,0]$.
\begin{equation}
\mbox{
\scriptsize{
$\begin{array}{|c|cccccc}
\dots&\dots&\dots&\dots&\dots&\dots&\dots\\
7&[0,0,0,0,7]&0&0&0&0&0\\
6&[0,0,0,0,6]&0&0&0&0&0\\
5&[0,0,0,0,5]&0&0&0&0&0\\
4&[0,0,0,0,4]&0&0&0&0&0\\
3&[0,0,0,0,3]&0&0&0&0&0\\
2&[0,0,0,0,2]&0&0&0&0&0\\
1&[0,0,0,0,1]&0&0&0&0&0\\
0&[0,0,0,0,0]&0&0&0&0&0\\\hline
&0&1&2&3&4&5
\end{array}$
}
}
\end{equation}

\section{D=6 N=1 YM}\label{S:rwerdk}
The goal of the section is to compute cohomology of the complex \ref{E:mxbdgdi} when dimension of the space-time is equal to $6$.

We define the space of pure spinors to be $\mathbf {P}^3\times\mathbf {P}^1$. Later we will give some arguments in support of this statement.

Introduce notations 
$T=\mathbb{C}^4,W=\mathbb{C}^2,V=\Lambda ^2[T],\mathbf {P}(T^*)=\mathbf {P}^3,\mathbf {P}(W^*)=\mathbf {P}^1$.
Then $H^0(\mathbf {P}^3,{\cal O}(1))=T,H^0(\mathbf {P}^1,{\cal O}(1))=W$.
Then Picard group of $\mathbf {P}^3\times \mathbf {P}^1$ is equal to $\mathbb {Z}+\mathbb {Z}$.
The sheaves ${\cal O}_{\mathbf {P}^3}(1)\boxtimes {\cal O}_{\mathbf {P}^1}$ and ${\cal O}_{\mathbf {P}^3}\boxtimes {\cal O}_{\mathbf {P}^1}(1)$ will be denoted by ${\cal O}(1,0)$ and ${\cal O}(0,1)$. They are generators of $Pic (\mathbf {P}^3\times \mathbf {P}^1)$.
We also have $H^0({\cal O}(1,1))=T\otimes W\overset {def}{=}S$.

In this section Greek indecies range from 1 to 4, Latin takes values 1,2. The vector space $S$ has basis $\theta_{\gamma a}$. The vector space $V=\Lambda ^2T$ has basis $A_{\gamma\delta}, \gamma<\delta$. We extend it to all values of index bu the rule $A_{\gamma\delta}=-A_{\delta\gamma}$. The space $V$ has coordinates $x^{\gamma\delta}$.

There are two Levi-Chevita totally skewsymmetric symbols $\epsilon^{ab}$ and $\epsilon^{\alpha\beta\gamma\delta}$. We have $\epsilon^{12}=1$ and $\epsilon^{1234}=1$
\begin{proposition}\label{P:rxcgsuj}
Consider a manifold $\mathbf {P}^3\times \mathbf {P}^1$ with a line bundle ${\cal O}(1,1)$.
Let $A^t$ be  a coordinated ring  of the embedding to $\mathbf {P}^7$ associated with this line bundle.
Then $A^t$ admits a resolution over symmetric algebra $\Sym(S)$, which has a form $A^t_i\leftarrow \Sym ^{i-2}(S)\otimes V\leftarrow \Sym ^{i-3}(S)\otimes S^*\leftarrow \Sym ^{i-4}(S)\otimes W$.
\end{proposition}
\begin{proof}
The proof will follow closely the proof of proposition 63 from \cite{MSch2}.

In order to construct the generating linear spaces of $i$-th free module of the minimal free resolution of $A^t$ over $\Sym (S)$, it is suffice to determine $\Tor _i^{\Sym (S)}(\mathbb{C},A^t)$.
The later groups are the cohomology groups of Koszul complex $A^t\otimes \Lambda[S]$.
If $\theta^{\gamma a}$ are generators of $\Lambda [S]$ and $\lambda^{\gamma a}$ are generators of $A^t$, then $d\theta ^{\gamma a}=\lambda^{\gamma a }$.
The algebra $A^t$ is a space of global sections of ${\cal C}=\bigoplus_{i\in \mathbb{Z}} {\cal O}(i,i)$.
We replace $A^t$ by ${\cal C}$ and retain the differential.
The arguments from proposition $63$ \cite{MSch2} show that the complex ${\cal C} \otimes \Lambda [S]$ splits into a direct sum of finite acyclic complexes $\bigoplus _{k \in \mathbb {Z}}(\bigoplus _{i\in \mathbb {Z}}{\cal O}(i,i)\otimes \Lambda ^{k-i}[S])$.
\begin{lemma}\label{L:ghsgfsd}
$H^p(\mathbf {P}^3\times \mathbf {P}^1,{\cal O}(-i,-i))$ is equal to zero unless $i\geq4, p=4$. In this case $H^4(\mathbf {P}^3\times \mathbf {P}^1,{\cal O}(-i,-i))=H^0(\mathbf {P}^3\times \mathbf {P}^1,{\cal O}(i-4,i-2))^*=\Sym^{i-4}(T)\otimes\Sym^{i-2}(W)$.
\end{lemma}
\begin{proof}
The proof will be given in section \ref{S:xdqde}
\end{proof}

Denote the $j$-th cohomology of the Koszul complex 
\begin{equation}
\dots \rightarrow \Lambda [S]\otimes A^t_{i-1}\rightarrow A^t_i=K(i)
\end{equation}
by $H^j(K(i))$.

Let $I$ be equal to $\bigoplus _{i\geq0}I_i=\bigoplus _{i\geq 0}\Sym ^i(T)\otimes \Sym ^{i+2}(W)$. It is $\Sym (S)$ module.
Let $M(i)$ be the Koszul complex of module $I$
\begin{equation}
M(i)=\dots \rightarrow \Lambda ^1[S]\otimes I_{i-1}\rightarrow I_i
\end{equation}
denote the $j$-th cohomology of $M(i)$ by $H^j(M(i))$.

\begin{proposition}\label{P:jwyb}
$H^n(K(k))=0,n\neq k,k-1$, $H^1(K(2))=\Lambda^2[T]$, $H^2(K(3))=T\otimes W$, $H^3(K(4))=\Sym^2(W)$ and trivially $H^0(K(0))=\mathbb{C}$. There is also identification
 \begin{equation}\label{PPLOk6}
\begin{split}
&H^k(K(k))=H^{3-k}(M(4-k))^*\\
&H^{k-1}(K(k))\cong H^{4-k}(M(4-k))^*
\end{split}
\end{equation}
Also $H^1(M(4))=\mathbb{C} $. All other cohomology of $M$ are equal to zero.
\end{proposition}
\begin{proof}
From spectral sequence arguments similar to proof of proposition 63  \cite{MSch2} we infer that \ref{PPLOk6} holds true. 
The cohomology groups $H^n(K(k))=0,n>k$ $H^n(M(k))=0,n>k$. With the aid of isomorphism \ref{PPLOk6} we conclude that $H^n(K(k))=0,n\neq k,k-1$ similarly $H^n(M(k))=0,n\neq k,k-1$ and $k$ should be in the range $0\leq k\leq 4$.

If $k=0$, then obviously $H^0(K(0))=\mathbb{C} $ and $H^n(K(0))=0,n>0$.
Therefore $H^1(M(4))=\mathbb{C} $ and $H^n(M(1))=0,n\neq 1$.

Suppose $k\neq 0$ then it is more or less obvious that $H^k(K(k))=0,H^n(K(k))\neq 0,n=k-1$.

Suppose $k=1$. Then by definition $H^0(K(1))=0$.

Suppose $k=2$. This is the first nontrivial case. We have a complex
\begin{equation}
K(2)=\Lambda^2[S]\rightarrow S\otimes S \rightarrow \Sym ^2T\otimes \Sym ^2W.
\end{equation}
With the aid of LiE we conclude that cohomology of this complex is equal to $V=\Lambda ^2T$.

Suppose $k=3$. We shall work with the complex
\begin{equation}
M(1)=S\otimes \Sym ^2W\rightarrow T\otimes \Sym ^3W
\end{equation}
The cohomology of this complex is equal to $S$. We conclude that cohomology $H^2(K(3))=S^*$.
The complex $M(0)$ has no differential and is equal to $\Sym ^2W$.
It implies that $H^3(K(4))=\Sym ^2W$.
\end{proof}

We can summarize the above computations in a statement that the structure of the minimal $\Sym(S)$ resolution  of $A^t$ is
\begin{equation}\label{PPLO17}
A^t\overset {d_0}{\leftarrow }\Sym (S)\overset {d_1}{\leftarrow }\Sym (S)\otimes V\overset {d_2}{\leftarrow }\Sym (S)\otimes S^*\overset {d_3^+}{\leftarrow }\Sym (S)\otimes \Sym ^2(W).
\end{equation}
It easy to determine the homomorphisms $d_i$ from \ref{PPLO17}. They are uniquely determined by $SL_4\times SL_2$-equivariance and homogeneity.

The map $d_1$ embeds generating space $V=\Lambda^2[T]\otimes \Lambda^2[W]$ into $\Sym^2(S)=\Sym^2(T\otimes W)$.

The map $d_2$ embeds $S^*=T^*\otimes W$ into $S\otimes V=T\otimes W\otimes \Lambda^2[T]$ via maps $T^*\cong \Lambda^3[T]\rightarrow T\otimes \Lambda^2[T]$ on $T^*$-component and identity on $W$-component.

The map $d_3^+$ maps $\Sym^2(W)$ into $S\otimes S^*=T\otimes W\otimes T^*\otimes W$ as $d_3^+(a\otimes a)=\sum_{i=1}^4 e_i\otimes a\otimes  e^i \otimes a$ ($e_i$ is a basis of $T$)
\end{proof}

In the course of the proof of proposition \ref{P:rxcgsuj} and proposition \ref{PPLOk6} we proved the following statement
\begin{proposition}
The module $I$ admits a resolution
\begin{equation}\label{PPLO22}
I^i\leftarrow \Sym ^2W\otimes \Sym ^i(S)\overset {d_3^-}{\leftarrow }S\otimes \Sym ^{i-1}(S)\overset {d_4}{\leftarrow }V\otimes\Sym ^{i-2}(S)\overset {d_5}{\leftarrow }\Sym ^{i-4}(S).
\end{equation}
\end{proposition}
The reader can reconstruct the values of $d^-_3,d_4,d_5$ following the proof the similar statement for $d_1,d_2,d^+_3$  of proposition \ref{P:rxcgsuj} as a guideline.
\begin{proposition}
The cohomology of the complex \ref{E:mxbdgdi} when D=6,N=1 are given in the table

\begin{equation}\label{T:nmxghs}
\mbox{
\scriptsize{
$\begin{array}{|c|cccccc}
\dots&\dots&\dots&\dots&\dots&\dots&\dots\\
8&[8,0,0,8]&0&[4,0,0,6]&0&0&0\\
7&[7,0,0,7]&0&[3,0,0,5]&0&0&0\\
6&[6,0,0,6]&0&[2,0,0,4]&0&0&0\\
5&[5,0,0,5]&0&[1,0,0,3]&0&0&0\\
4&[4,0,0,4]&0&[0,0,0,2]&0&0&0\\
3&[3,0,0,3]&0&0&0&0&0\\
2&[2,0,0,2]&0&0&0&0&0\\
1&[1,0,0,1]&0&0&0&0&0\\
0&[0,0,0,0]&0&0&0&0&0\\\hline
&0&1&2&3&4&5
\end{array}$
}
}
\end{equation}

\end{proposition}
\begin{proof}
We sew  the complexes \ref{PPLO17} and \ref{PPLO22} over $\Sym ^2W\otimes \Sym (S)$. The resulting complex will coincide with \ref{E:mxbdgdi}. The check could be performed by direct inspection. It is straightforward except identification of  $d_3$ form \ref{E:mxbdgdi}  with $d^+_3\circ d^-_3$ from \ref{PPLO17} and \ref{PPLO22}. It could be done with a help of LiE program. The cohomology of the sewed complex is a direct sum $A^t$ and $I$ . This is reflected in the table \ref{T:nmxghs}.
\end{proof}
%

%
The cohomology of algebra $L$ carry a structure of homogeneous A$_{\infty}$-algebra. It means that the algebra is bigraded. The first grading  is the cohomological, the second is the  internal grading (coming from the grading of the algebra $L$). By definition the cohomological degree of a structure map $\mu_n$ is equal to $2-n$. The internal degree of the same map is equal to zero (the reader can consult \cite{MSch2} for details). 


In our case the bidegree of $A^t_i$ is equal to $(2i,i)$ and of $I_i$ is $(6+2i,4+i)$.

\begin{proposition}\label{P:qwfxjk}
The only possible operations in A$_{\infty}$-algebra $A=A^t+I$ are

{\bf binary:} Multiplication in $A^t$, $A^t$-module structure on $I$.

{\bf ternary:} A linear map $c:A^{t\otimes 4}\rightarrow I$. It is Hochschild 4-cocycle.( Recall that it is a linear map which satisfies an equation
\begin{equation}
a_0c(a_1,\dots,a_4)+\sum_{i=0}^3(-1)^{i+1} c(a_0,\dots,a_ia_{i+1},\dots,a_4)-c(a_0,\dots,a_3)c_4=0
\end{equation}
\end{proposition}
Our next task shall be an identification of the cocycle $c$.

The element $c$ belongs to a group $Hom(A^{t\otimes 4},I)$, which is a member of Hochschild complex with coefficients in bimodule $I$ (see \cite{MSch2} for definition). It is rather hard to compute cohomology of the Hochschild complex directly in general . We are going to take advantage of the fact that $A^t$ is a Koszul algebra (see \cite{Bezr}).
%
We can take advantage of proposition \ref{P:vdsyh} and identify cohomology $HH(A^t,I)$ with cohomology of the complex $(I\otimes A^{t!},d)$.
The algebra $A^{t!}$, being Koszul dual to commutative algebra $C$ has commutators as relations. Hence $A^{t!}$ is a universal enveloping of a Lie algebra $\mathfrak{c}^!$. Since the differential $d$ in $I\otimes A^{t!}$ is a commutator and $\theta_{\alpha}$ are primitive, then $I\otimes \mathfrak{c}^!$ is a subcomplex of $I\otimes A^{t!}$. By Poincare-Poincare-Birkhoff-Witt theorem $A^{t!}=\bigoplus_{i\geq 0}\Sym^i(\mathfrak{c}^!)$. The decomposition is a direct sum of $\mathfrak{c}^!$-modules. From this we conclude that $I\otimes \mathfrak{c}^!$ is a direct summand of $I\otimes A^{t!}$. Denote $I\otimes A^{t!}(n)$ the subcomplex of elements of internal degree $n$. Define similarly $I\otimes \mathfrak{c}^!(n)$.

The Lie algebra $\mathfrak{c}^!$ was introduced for the first time in mathematical literature in \cite{DF}.

 The Bar-dual to the A$_{\infty}$ algebra $A$ is a universal enveloping of $L$. From this we conclude that the cocycle $c$ can be represented by an element of $I\otimes \mathfrak{c}^!$ .
A more accurate computation of bidegree shows that the cocycle must be an element of $I_0\otimes \mathfrak{c}^!_4$

The low-dimensional components of $\mathfrak{c}^!_n$ can be computed explicitly
\begin{equation}
\begin{split}
&\mathfrak{c}^!_1=[1,0,0,1]\\
&\mathfrak{c}^!_2=[0,1,0,0]\\
&\mathfrak{c}^!_3=[0,0,1,1]\\
&\mathfrak{c}^!_4=[0,0,0,2]+[1,0,1,0]\\
&\dots
\end{split}
\end{equation}
We have an obvious $SL_4\times SL_2$-invariant element in $I_0\otimes \mathfrak{c}^!_4$ which is a generator of one-dimensional $SL_4\times SL_2$ vector space in $[0,0,0,2] \otimes [0,0,0,2]\subset I_0\otimes \mathfrak{c}^!_4$. By the existence theorem (proposition \ref{P:qwfxjk}) this element must be nontrivial $d$-cocycle.
\subsection{Geometric interpretation of the cocycle $c$ }
We would like to give a more algebro-geometric description of  cocycle $c$. We take advantage of sheafification procedure which was of great help to us in computation of cohomology of complexes $K(i)$ and $M(i)$. We replace a $A^t$ module $I$ by (infinite dimensional) vector bundle ${\cal I}=\bigoplus_{i\in \mathbb{Z}}{\cal O}(i,i+2)$. In the complex of sheaves  ${\cal I}\otimes \mathfrak{c}^!$ we choose a direct summand
\begin{equation}\label{E:qerhc}
\mathfrak{c}^!_1{\cal O}(-3,-1)\rightarrow \mathfrak{c}^!_2{\cal O}(-2,0)\rightarrow\mathfrak{c}^!_3{\cal O}(-1,1)\rightarrow\mathfrak{c}^!_4{\cal O}(0,2)\rightarrow\mathfrak{c}^!_5{\cal O}(1,3)\rightarrow \dots
\end{equation}

From our computation of cohomology of sheaves ${\cal O}(i,i+2), i\geq -3$ we know that they have no higher cohomology. Hence the hypercohomology of \ref{E:qerhc} coincide with cohomology of $I\otimes \mathfrak{c}^!(0)$. 

In such situation local to global method works particularly well. In the local approach we  specialize the complex \ref{E:qerhc} to a point in $\mathbf{P}^3\times \mathbf{P}^1$. Then we  compute  cohomology of the  reduced complex, obtained in the result of specialization. After that we return back to the global picture, considering a sheaf over $\mathbf{P}^3\times \mathbf{P}^1$, whose fibers are the computed cohomology.

Any $\theta\in \mathfrak{c}^!_1$ such that
\begin{equation}\label{E:uisgw}
\{\theta,\theta\}=0
\end{equation}
 defines a differential $d_\theta(a)=\{\theta,a\}$. Denote $\theta_{\alpha}$ is a basis of $\mathfrak{c}^!_1$ and $A_i$ is a basis of $\mathfrak{c}^!_2$. Because of the general relation $\{\theta_{\alpha},\theta_{\beta}\}=\Gamma_{\alpha\beta}^iA_i$ in algebra $\mathfrak{c}^!$ which in th present context becomes $\{\theta_{\alpha a},\theta_{\beta b}\}=\epsilon{ab}A_{\alpha\beta}$ we conclude that  $\theta$  defines a differential iff $\theta$ is pure. The cohomology of $d_\theta$ is  the deformation of a solution of equation \ref{E:uisgw}.

Denote $Y$ the total space of the line bundle ${\cal O}(-1,-1)$ over $\mathbf{P}^3\times\mathbf{P}^1$. It is obvious that any nonzero solution $\theta$ of equation \ref{E:qerhc} can be represented by the  point $\theta$ in $Y$ away of the zero section.
\begin{proposition}
The cohomology of the complex $(\mathfrak{c}^!,d_{\theta})$ coincide with the tangent space of $Y$ to the point $\theta$.
\end{proposition}
\begin{proof}
A rigorous proof of this statement (vanishing of higher cohomology ) in a broader context will be given in \cite{MSch5}.
\end{proof}

The cohomology groups are unchanged upon rescaling  $\theta$. Thus the cohomology defines a vector bundle over $\mathbf{P}^3\times\mathbf{P}^1$. It is easy to identify the bundle

Denote the restriction of the tangent bundle to $Y$ to the zero section by $\widetilde{T}$. Then the sheaf of cohomology of \ref{E:qerhc} is equal to $\widetilde{T}(-3,-1)$.

 There is a short exact sequence
\begin{equation}\label{E:iocnswg}
0\rightarrow {\cal O}(-1,-1)\rightarrow\widetilde{T}\rightarrow T(-1,-1)\rightarrow 0
\end{equation}
where $T$ is the tangent bundle to $\mathbf{P}^3\times\mathbf{P}^1$
\begin{proposition}
The cohomology of the sheaf $\widetilde{T}(-3,-1)$ is nonzero only in dimension three where it is one-dimensional.
\end{proposition}
\begin{proof}
Use long exact sequence associated with \ref{E:iocnswg}, twisted on ${\cal O}(-3,-1)$
\end{proof}
%
%

Denote  $\nu \in \Omega^{0,3}\tilde{T}(-3,-1)$ a nontrivial cocycle representing a generator of $H^3(\tilde{T}(-3,-1))$.
It is easy to see that there is a projection of complexes
\begin{equation}
Hom(A^{t\otimes n},I)\rightarrow I\otimes A^{t!}_n
\end{equation}
It is restriction on $A^t_1$ with further factorization.

The element $c$ was originally introduced as an element of $Hom(A^{t\otimes 4},I)$. Our goal is to recover it from $\nu$.

We sheafify $Hom(A^{t\otimes n},I)$ by replacing $I$ by ${\cal I}$.

We will be interested in the direct summand
\begin{equation}
Hom(A^{t\otimes 0},{\cal O}(-4,-2))\rightarrow Hom(A^{t\otimes 1},{\cal O}(-3,-1))\rightarrow \dots
\end{equation}
The cohomology groups of the individual sheaves in this complex have very simple cohomology: $Hom(A^{t\otimes i},{\cal O}(-4+i,-2+i))$ have only zero cohomology for $i\geq4$, no cohomology at all if $1\leq i\leq3$ and fourth cohomology for $i=0$. 

We can use a transgression to construct an element $c$. 

Denote ${\cal L}$ one of the line bundles ${\cal O}(-3,-1),{\cal O}(-2,0),{\cal O}(-1,1)$. From acyclicity there is an operator $h_{\cal L}:\Omega^{0,p}\otimes {\cal L}\rightarrow \Omega^{0,p-1}\otimes {\cal L}$ which satisfies $\{h_{\cal L},\dbar\}=id$. The operator $h_{\cal L}$ can be represented by an integral kernel $G_{\cal L}$. The object $G_{\cal L}\in \Omega^{4,3}_{\mathbf{P}^3\times\mathbf{P}^1\times\mathbf{P}^3\times\mathbf{P}^1}\otimes {\cal L}\boxtimes {\cal L}^{-1}$. The form $G_{\cal L}$ has a singularity on the diagonal $\Delta \subset \mathbf{P}^3\times\mathbf{P}^1\times\mathbf{P}^3\times\mathbf{P}^1$: if we make a blowup of the diagonal then the corresponding divisor $D$ locally will be defined by the equation $z=0$.  The pullback of the form $G_{\cal L}$ on blowup satisfies the property that $zG_{\cal L}$ can be extended on the divisor.
There is an inclusion $\tilde{T}(-3,-1)\subset \mathfrak{c}_1^!(-3,-1)$. This we have an embedding of Dolbeault cochains.
\begin{proposition}
The desired cocycle $c$ is $dh_{{\cal O}(-1,1)}dh_{{\cal O}(-2,0)}dh_{{\cal O}(-3,-1)} \nu$, where $d$ is the Hochschild differential.
\end{proposition}
\begin{proof}
By construction.
\end{proof}

Instead of trying to find explicitly $G_{\cal L}$ one can work with local model of algebra $A$. This model will be a direct sum
\begin{equation}
{\cal A}=\bigoplus_{i\geq 0}\Omega^{0,\bullet}{\cal O}(i,i)+\bigoplus_{i\geq -3}\Omega^{0,\bullet}{\cal O}(i,i+2)
\end{equation}
The second direct summand is the ideal with zero multiplication. The algebra ${\cal A}$ is equipped with differential $\dbar+\nu$. The element $\nu$ is interpreted as differentiation
\begin{equation}
\nu:\Omega^{0,p}{\cal O}(i,i)\rightarrow \Omega^{0,p+3}{\cal O}(i-3,i-1)
\end{equation}
\begin{proposition}
The algebra ${\cal A}$ is quasiisomorphic to $A$.
\end{proposition}
\begin{proof}
By construction.
\end{proof}

One can apply Berkovits construction to the algebra ${\cal A}$:
\begin{definition}
On the tensor product ${\cal A}\otimes \Lambda[S]$ define a differential $d=\lambda^{\gamma a}\pr{\theta_{\gamma a}}$. and the total differential $\dbar+\nu+d$
\end{definition}
\begin{proposition}
The algebra $({\cal A}\otimes \Lambda[S],\dbar+\nu+d)$ is quasiisomorphic to Cartan-Chevalle complex of Lie algebra $YM$.
\end{proposition}
\begin{proof}
We present only sketch of the proof.

We make computation of the cohomology using spectral sequence arguments. This is done by means of computations of cohomology of differentials $\dbar, d, \nu$. The first two differentials give almost right cohomology with difference being the auxiliary fields. Later are killed by $\nu$.

The presence of right higher multiplications  is made by direct inspection.
\end{proof}

The superspace formulation of $N=1,D=6$ YM gives an  explanation of some higher multiplication  in $A$. We  work in superspace formulation of the theory reduced to a point.

We have a space  $\mathbb{C}^{6|8}=\Lambda^2[T]+\Pi(T\otimes W)$. It carries a nonintegrable distribution $G$ spanned by vector fields $\tau_{\gamma a}=\pr{\theta_{\gamma a}}+\theta_{\delta a}\pr{x^{\gamma \delta}}$.

It will be useful for us to understand the structure of the ideal  $\mathfrak{i}=\bigoplus_{i\geq 2}\mathfrak{c}^!$. 
The following proposition is byproduct of the proof of the proposition \ref{P:jwyb}:
\begin{proposition}\label{P:kdhdg}
The algebra $\mathfrak{i}$ is a free Lie algebra generated by $V=\Lambda^2[T],S=T\otimes W, \Sym^2(W)$. The supersymmetry transformations $\mathfrak{c}^!$ act by outer derivations of $\mathfrak{i}$.
\end{proposition}
\begin{proof}
To compute generators and relations of $\mathfrak{i}$ due to proposition \ref{E:mkdgaf} we need to compute cohomology of the complex $A^t\otimes \Lambda[S]$. The cohomology of this complex where computed in proposition \ref{P:jwyb}. Since the auxiliary dimension of all nontrivial cohomology groups is equal to one then by proposition \ref{P:lswcd} the Lie algebra $\mathfrak{i}$ is free.
\end{proof}
\begin{remark}
We may interpret $V,S, \Sym^2(W)$ as a connection, spinor field , and auxiliary quaternionic field
\end{remark}

To transform a constrained partial $\mathbb{C}^6$-invariant connection ${\cal D}_{\gamma a}$ into a solution of MC equation for the algebra $A\otimes \Lambda[S]$ is to represent ${\cal D}_{\gamma a}=\tau_{\gamma a}+A_{\gamma a}$.  We assign to $A_{\gamma a}$ an element $\lambda^{\gamma a}A_{\gamma a}$.
\begin{proposition}
An element $m=\lambda^{\gamma a}A_{\gamma a}\in A^t\otimes \Lambda[S]\otimes Mat_n$ constructed from constrained connections is a solution of MC equation . Any constrained connection can be obtained this way.
\end{proposition}
\begin{proof}
This is a simple corollary of \ref{P:kdhdg}.
\end{proof}

We know that the subalgebra $\mathfrak{i}\subset \mathfrak{c}^!$ has three dimensional generating space  $W$ in the fourth graded component. In terms of generators $w^{\alpha i}$ of $\mathfrak{c}^!$ such element could be written in the form
\begin{equation}\label{E:fqwqc}
E_i^j= \epsilon^{\gamma\delta\gamma'\delta'}\epsilon^{mj}\epsilon^{kl}\{w_{\gamma i},\{w_{\delta m},\{w_{\gamma' k},w_{\delta' l}\}\}\}
\end{equation}

On the level of the superfields we have the following fields
\begin{equation}
\begin{split}
&\nabla_{\alpha\beta}=\epsilon^{ij}\{{\cal D}_{\alpha i}, {\cal D}_{\beta j}\}\\
&\Lambda^{\delta j}=\epsilon^{\alpha\beta \gamma\delta}\epsilon^{mj}[\nabla_{\alpha\beta},{\cal D}_{\gamma m}]
\end{split}
\end{equation}

Then the auxiliary field can be found as
\begin{equation}
{\cal E}^i_j=\{{\cal D}_{\alpha j},\Lambda^{\alpha i}\}
\end{equation}
Expressing ${\cal E}^i_j$ in terms of defining partial connections ${\cal D}_{\alpha a}$  we arrive to  equation \ref{E:fqwqc}
\section{D=4 N=1 YM}

Denote $S=W_{\mathbf{ l}}+W_{\mathbf {r}}$, $dim(W_{\mathbf {l}})=dim(W_{\mathbf{r}})=2$

\begin{proposition}
The cohomology $A$ of the complex \ref{E:mxbdgdi} adapted to the setting of $D=4,N=1$ YM theory are  tabulated   below:

\begin{equation}\label{T:xcdk}
\mbox{
\scriptsize{
$\begin{array}{|c|cccccc}
\dots&\dots&\dots&\dots&\dots&\dots&\dots\\
7&[7,0]+[0,7]&0&0&[1,0]+[0,1]&0&0\\
6&[6,0]+[0,6]&0&0&[0,0]+[0,0]&0&0\\
5&[5,0]+[0,5]&0&0&0&0&0\\
4&[4,0]+[0,4]&0&[0,0]&0&0&0\\
3&[3,0]+[0,3]&0&0&0&0&0\\
2&[2,0]+[0,2]&0&0&0&0&0\\
1&[1,0]+[0,1]&0&0&0&0&0\\
0&[0,0]&0&0&0&0&0\\\hline
&0&1&2&3&4&5
\end{array}$
}
}
\end{equation}
The elements of the $i$-th row in the zeroth  column form a group $A_{2i,i}$. The one dimensional space in the second column is denoted by $<c>=A_{6,4}$. The group in the third column, $i$-th row is denoted by $A_{2i-3,i}$ $i\geq 6$. In the subscript we have topological and homogeneous gradings.
\end{proposition}
\begin{proof}

The space $V=W_{\bold l}\otimes W_{\bold r}$ has coordinates $x_{\alpha \dt{\beta}}$. The space $\Pi W_{\bold l}$ has coordinates $\theta_{\alpha}$, the space $\Pi W_{\bold r}$ has  $\theta_{\dt{\beta}}$, $\alpha, \dt{\beta}=1,2$. Denote $\lambda^{\alpha}, \lambda^{\dt{\beta}}$ generators of $A^t$.

Let us consider a minimal resolution of the $\Sym(S)$-module $A^t=\Sym(S)/W_{\bold l}\otimes W_{\bold r}$.
\begin{lemma}\label{L:hddxt}
The minimal resolution of $A^t$ considered as $\Sym(S)$-module has the form
\begin{equation}\label{E;abdgs}
A^{ti}\leftarrow \Sym^i(S)\overset{d_0}{\leftarrow} \Sym^{i-2}(S)\otimes W_{\bold l}\otimes W_{\bold r}\overset{d_1}{\leftarrow} \Sym^{i-3}(S)\otimes S\overset{d^+_2}{\leftarrow} \Sym^{i-4}(S)
\end{equation}
\end{lemma}
\begin{proof}
The main observation is that  the ring and the module in this construction admit an action of $\mathbb{Z}_2\ltimes SL_2\times SL_2$, where the generator of $\mathbb{Z}_2$ swaps the $(a,b)\in SL_2\times SL_2$. Since the minimal resolution in graded case could be constructed in $\mathbb{Z}_2\ltimes SL_2\times SL_2$-equivariant fashion we should expect that all differential in \ref{E;abdgs} are maps of $\mathbb{Z}_2\ltimes SL_2\times SL_2$-representations. If we prove that the modules in \ref{E;abdgs} are the same as in minimal resolution then the equivariance condition completely fixes the differential. 

To compute the the degrees and representation content of the generators we compute cohomology of $A^t\otimes \Lambda[S]$. This is completely analogous to the computation given in section \ref{S:rwerdk} for the group $SL_4\times SL_2$. Instead of the proof we list the generators:
\begin{equation}
\begin{split}
&<\lambda^{\alpha}\theta^{\dt{\beta}}>\\
&<\lambda^{\alpha}\theta^{\dt{1}}\theta^{\dt{2}},\lambda^{\dt{\alpha}}\theta^{1}\theta^{2}>\\
&<\lambda^{1}\theta^{2}\theta^{\dt{1}}\theta^{\dt{2}}-\lambda^{2}\theta^{1}\theta^{\dt{1}}\theta^{\dt{2}}>
\end{split}
\end{equation}
\end{proof}

The next goal is a computation of cohomology of the complex
\begin{equation}\label{E:afdejf}
\Sym^i(S)\overset{d^-_2}{\leftarrow}\Sym^{i-1}(S)\otimes S\overset{d_3}{\leftarrow}\Sym^{i-2}(S)\otimes W_{\bold l}\otimes W_{\bold r}\overset{d_4}{\leftarrow} \Sym^{i-4}(S)
\end{equation}
equipped with the only possible  nontrivial, $\mathbb{Z}_2\ltimes SL_2\times SL_2$-equivariant $\Sym(S)$-linear  differential. 

\begin{lemma}
The cohomology of the complex \ref{E:afdejf} is equal to $\mathbb{C}$ in dimension zero and $I=\Sym(W_{\bold l})+\Sym(W_{\bold r})$ in dimension one.
\end{lemma}
\begin{proof}
In our computation we are going to compare \ref{E:afdejf} with Koszul complex $\Sym(S)\otimes \Lambda[S]$. In more details the $i$-th graded component of the  later is equal to
\begin{equation}
\Sym^i(S)\overset{\mu_0}{\leftarrow}\Sym^{i-1}(S)\otimes S\overset{\mu_1}{\leftarrow}\Sym^{i-2}(S)\otimes(W_{\bold l}\otimes W_{\bold r}+\mathbb{C}^2)\overset{\mu_2}{\leftarrow} \Sym^{i-3}(S)\otimes S \overset{\mu_3}{\leftarrow}\Sym^{i-4}(S)
\end{equation}
 It  is acyclic for $i>0$. 

Returning to the complex \ref{E:afdejf} we observe that the module $\Sym(S)\otimes W_{\bold l}\otimes W_{\bold r} $ has no torsion, hence $d_4$ is an embedding.

By  construction $d^-_2=\mu_0$, hence
\begin{equation}
\Ker d^-_2=\Sym(S)\otimes(W_{\bold l}+W_{\bold r}+\mathbb{C}^2)/\Imm \mu_2)
\end{equation}
 Thus the complex \ref{E:afdejf} is quasiisomorphic to
\begin{equation}\label{E:jhagwr}
\Sym(S)\otimes(W_{\bold l}\otimes W_{\bold r}+\mathbb{C}^2)/\Imm \mu_2 \overset{f}{\leftarrow} \Sym(S)\otimes W_{\bold l}\otimes W_{\bold r}/\Sym(S)
\end{equation}
The map $f$ is induced by inclusion $\Sym(S)\otimes W_{\bold l}\otimes W_{\bold r} $ into $\Sym(S)\otimes( W_{\bold l}\otimes W_{\bold r} +\mathbb{C}^2)$. Denote the image of $\Sym(S)$(it is a submodule of the module $\Sym(S)\otimes W_{\bold l}\otimes W_{\bold r}$ ) in $\Sym(S)\otimes( W_{\bold l}\otimes W_{\bold r} +\mathbb{C}^2)$ by $X$.  We have an identity
\begin{equation}
\epsilon^{\alpha \beta}\epsilon^{{\dt{\alpha}}{\dt{\beta}}} w_{\alpha}\wedge w_{\dt{\alpha}}\otimes w_{\beta} w_{\dt{\beta}}=\frac{1}{2}\mu_2(\epsilon^{\alpha \beta}\epsilon^{{\dt{\alpha}}{\dt{\beta}}} w_{\alpha} \wedge w_{\beta}\wedge w_{\dt{\alpha}}  \otimes w_{\dt{\beta}})
\end{equation}
which insures that $X\subset \Imm \mu_2$ and the map \ref{E:jhagwr} is correctly defined. The map $f$ is injective. This could be proved by explicitly solving equation $\mu_2x=a$, where $a$ has $\Sym(S)\otimes \mathbb{C}^2$-component in $\Sym(S)\otimes(W_{\bold l}+W_{\bold r}+\mathbb{C}^2)$ equal to zero. 

The space $\mathbb{C}^2$ has a basis $a_{\bold l},a_{\bold r}$ which is invariant with respect to $SL_2\times SL_2$ action. The elements are swapped by the generator of $\mathbb{Z}_2$.
Thus the cohomology of the complex \ref{E:afdejf} is equal to $\Sym^i(S)\otimes \mathbb{C}^2/\Sym^i(S)\otimes \mathbb{C}^2\cap Im p\mu_2$( $p$ is a projection on $\mathbb{C}^2\otimes\Sym(S)$ component) for $i>0$. From the form of $p\mu_2$ one can see in straightforward manner that $Im p\mu_2$ is generated by $<a_{\bold l}>\otimes W_{\bold r}+<a_{\bold r}>\otimes W_{\bold l}$. Thus the $A^t$-module structure on the cokernel of map $f$ can be described as follows: there are two projections $p_{\alpha}:A^t\rightarrow \Sym{W_{\alpha}}$ $\alpha={\bold l},{\bold r}$. The  module  is $\Sym(W_{\bold r})+\Sym(W_{\bold r})$. Only first cohomology of \ref{E:afdejf} appear nonzero on the homogeneous degree $i>0$.
\end{proof}

%

The last step is a computation of the cohomology of the complex \ref{E:mxbdgdi}. As in previous sections we sew the complexes \ref{E;abdgs}, \ref{E:afdejf} together, identifying the last term of the former with the fist term of the later. It not hard to see that such identification results in the complex \ref{E:mxbdgdi}.
From this we deduce the table \ref{T:xcdk}.


\end{proof}
\begin{proposition}
The algebra $A$ has the following nonzero operations: $\mu_2,\mu_3,\mu_4,\mu_5$.
\end{proposition}
\begin{proof}
We use a standard trick of enumeration all possible multiplications compatible with cohomological and homogeneous gradings.
As a result of simple arithmetic we get the answer: only $\mu_2,\mu_3,\mu_4,\mu_5$ are possible. We are left to prove that they all are present.

There is no questions about $\mu_2$. The operation $\mu_4$ if exists must be degenerate: it nontrivially maps only $A_{2,1}^{\otimes4}\rightarrow A_{6,4}=<c>$. This map must be nontrivial, otherwise we would get a contradiction with proposition \ref{P:ghsfa}.

If the map $\mu_3$ is not zero then it transforms $A^t\otimes A^t\cong A^t\otimes A^t \otimes <c>\rightarrow I$. We denote the compound map by $\psi$. Since the action of $A^t$ on $c$ is trivial we conclude  $\psi$ must be a Hochschild $A^t$ two-cocycle with values in $I$. One however must use caution, because the bimodule structure of $I$ is not standard. The left action of $A^t$ is standard, the right action is through augmentation. The reason is that the ordinary multiplication of element $c$ with $I$ is trivial. It is not hard to identify the cocycle  $\psi$ in the complex $A^{t!}\otimes I$ which computes such cohomology. We leave this task to the reader as an exercises and only point out that it is an element of $A^{t!}_2\otimes A_{9,6}\subset A^{t!}\otimes I$.

The maps $\mu=\sum_i\mu_i$ satisfy a quadratic relation. It  was already used in the previous paragraph in assertion that $\psi$ is a cocycle. The relation in question  has a form of bracket $[\mu,\mu]=0$ if we interpret $\mu$ as an element of some infinite-dimensional Lie algebra (see \cite{MSch2} for discussion and references ). The commutator is obtained as linear combination with signs of substitutions of $\mu_i$ for the arguments of $\mu_j$ and substitution of $\mu_j$ into $\mu_i$.

Having this interpretation of the relation we must have $[\mu_2,\mu_5]$  proportional to $[\mu_3,\mu_4]$. Thus  if $\mu_3\neq0$ then $[\mu_3,\mu_4]\neq0$. If we assume that $\mu_3=0$ then then it automatically implies that $[\mu_2,\mu_5]=0$. The last equation   is the same thing as equation   of  Hochschild 5-cocycle for $\mu_5$ of $A^t$ with values in $I$. 
This cocycle can be represented as $\mathbb{Z}_2\ltimes SL_2\times SL_2$ invariant element of $\mathfrak{c}_5^!\otimes A_{9,6}$. From proposition \ref{P:qhxh}, we see that $\mathfrak{c}_5^!$ is spanned by commutators $[v,s]$, $v\in V$, $s\in S$  and contains no $\mathbb{Z}_2\ltimes SL_2\times SL_2$ invariant elements.
\end{proof}

\subsection{Geometric interpretation}
Our next task will be to understand the structure of A$_{\infty}$-algebra on $A$ and interpret it from the point of view of superspace formalism.

The algebra $A^t$ is a direct sum of two Koszul algebras $\Sym(W_{\bold l})+\Sym(W_{\bold r})$. According to
\cite{Qalg} the operation of a direct sum preserves Koszul property. Koszul duality transforms a direct sum of two commutative algebras into a free product of Lie algebras. Thus the Koszul dual $\mathfrak{c}^!$ of $A$ is a universal enveloping of a free product of two abelian (odd) Lie algebras $W_{\bold l}^*\cong W_{\bold l}$ and $W_{\bold r}^*\cong W_{\bold r}$. In complete analogy with proposition \ref{P:kdhdg} we can state a proposition based on computation of cohomology of the complex $A^t\otimes \Lambda[S]$.
\begin{proposition}\label{P:qhxh}
The subalgebra $\mathfrak{i}$ of lie algebra $\mathfrak{c}^!$ is free. It is generated by $V=W_{\bold l}\otimes W_{\bold r}$ in degree 2, $S=W_{\bold l}+W_{\bold r}$ in degree 3, $E=\mathbb{C}$ in degree 4. The algebra $A^t\otimes \Lambda[S]$ is Bar-dual to $\mathfrak{i}$.
\end{proposition}
\begin{proof}
By lemma  \ref{L:hddxt} all nontrivial cocycles have auxiliary degree equal to one. Then we use
We use proposition \ref{E:mkdgaf} in conjunction with \ref{P:lswcd}. The statement about Bar-duality is a corollary of \ref{E:mkdgaf}.
\end{proof}
\begin{remark}
We interpret $V,S,E$ as gauge , spinor, auxiliary scalar fields
\end{remark}

The superspace formulation of $N=1,D=4$ YM offers en explanation of some higher multiplication  in $A$. As in section \ref{S:secsl2} we are going to work in superspace formulation of the theory reduced to a point.

We have a space  $\mathbb{C}^{4|4}=W_l\otimes W_r+\Pi(W_l+ W_r)$. It carries a nonintegrable distribution $G$ spanned by vector fields $\tau_{\alpha}=\pr{\theta_{\alpha}}+\epsilon^{\dt{\beta}\dt{\gamma}}\theta_{\dt{\gamma}}\pr{x_{\alpha \dt{\beta}}}$,  $\tau_{\dt{\alpha}}=\pr{\theta_{\dt{\alpha}}}+\epsilon^{\beta\gamma}\theta_{\gamma}\pr{x_{\beta \dt{\alpha}}}$.

A simple way to transform a constrained partial $\mathbb{C}^4$-invariant connection ${\cal D}_{\alpha}$ into a solution of MC equation for the algebra $A\otimes \Lambda[S]$ is to represent ${\cal D}_{\alpha}=\tau_{\alpha}+A_{\alpha}$ and ${\cal D}_{\dt{\alpha}}=\tau_{\dt{\alpha}}+A_{\dt{\alpha}}$. We assign to $A_{\alpha},A_{\dt{\beta}}$ and element $\lambda^{\alpha}A_{\alpha}+\lambda^{\dt{\beta}}A_{\dt{\beta}}$.
\begin{proposition}
An element $m=\lambda^{\alpha}A_{\alpha}+\lambda^{\dt{\beta}}A_{\dt{\beta}}\in A^t\otimes \Lambda[S]\otimes Mat_n$ constructed from constrained connections is a solution of MC equation . Any constrained connection can be obtained this way.
\end{proposition}
\begin{proof}
This is a simple corollary of \ref{P:qhxh}.
\end{proof}

We know that the subalgebra $\mathfrak{i}\subset \mathfrak{c}^!$ has generator $E$ in the fourth graded component. In terms of generators $w^{\alpha}, w^{\dt{\beta}}$ of $\mathfrak{c}^!$ such element could be written in the form
\begin{equation}\label{E:cqad}
E=\epsilon^{\alpha \beta} \epsilon^{\dt{\alpha}\dt{\beta}} \{w_{\dt{\alpha}},\{w_{\beta},\{w_{\alpha},w_{\dt{\beta}}\}\}\}
\end{equation}

On the level of the superfields we have the following fields
\begin{equation}
\begin{split}
&\nabla_{\alpha\dt{\beta}}=\{{\cal D}_{\alpha}, {\cal D}_{\dt{\beta}}\}\\
&\Lambda_{\alpha}=\epsilon^{\dt{\beta}\dt{\gamma}}[\nabla_{\alpha\dt{\beta}},{\cal D}_{\dt{\gamma}}]\\
&\Lambda_{\dt{\alpha}}=\epsilon^{\beta\gamma}[\nabla_{\beta\dt{\alpha}},{\cal D}_{\gamma }]
\end{split}
\end{equation}

Then the auxiliary field can be found as
\begin{equation}
{\cal E}=\epsilon^{\alpha\beta}\{{\cal D}_{\alpha},\Lambda_{\beta}\}
\end{equation}
Expressing ${\cal E}$ in terms of defining partial connections ${\cal D}_{\alpha}$, ${\cal D}_{\dt{\alpha}}$ we arrive to  equation \ref{E:cqad}
\section{D=3,N=1 YM}\label{S:secsl2}
Denote $W$ a two-dimensional vector space. In this section Greek indecies take values 1,2.
We start this section with definition of a certain linear map. There is a linear  equivariant identification $\Sym^2(W)\rightarrow \mathfrak{sl}_2$, $w_1\otimes w_2\rightarrow \{w_1,w_2\}$. There is also an action $\mathfrak{sl}_2\otimes W\rightarrow W$, $a\otimes w\rightarrow [a,w]\overset{def}{=}-[w,a]$. Define a map
\begin{equation}\label{E:xdgsy}
\mu_5:W^{\otimes 5}\rightarrow W
\end{equation}
as $w_1\otimes \dots \otimes w_5\rightarrow [\{w_1,w_2\},[\{w_3,w_4\},w_5]]$.
There are several way (up to a constant) to defined such a map. One of them is $\mu_5(w_1\otimes \dots \otimes w_5)=[w_1,[w_2,[w_3,\{w_4,w_5\}]]]$. 
\begin{proposition}
Define a A$_{\infty}$ homogeneous algebra which is a direct sum of spaces $A_{0,0}=\mathbb{C}, A_{2,1}=W,A_{7,5}=W, A_{9,6}=\mathbb{C}$. (The first grading is topological, the second is homogeneous).There is a pairing on $A$ of degree zero with values in $A_{9.6}$.  All higher  multiplications are equal to zero, except a quintic multiplication which coincides with \ref{E:xdgsy}. 
This A$_{\infty}$ algebra is a Bar-dual to supersymmetry algebra $L$ in three dimensions.

\end{proposition}
\begin{proof}
The maps $d_j(i)$ are defined in the table below. In it we tabulated the first entries of the complex \ref{E:mxbdgdi} specialized to the setting of the present section. We denote irreducible $\mathfrak{sl}_2$ representation in $\Sym^i(W)$ of highest weight $i$ by $[i]$.
\begin{equation}\label{E:fgojkxcz}
\mbox{
\scriptsize{
\setlength{\unitlength}{3947sp}
\begin{picture}(6312,4494)(226,-4261)
\thinlines
\put(1126,-1186){\vector(-1, 0){500}}
\put(1126,-1561){\vector(-1, 0){500}}
\put(1126,-1936){\vector(-1, 0){500}}
\put(2600,-1561){\vector(-1, 0){500}}
\put(2600,-1936){\vector(-1, 0){500}}
\put(2600,-2311){\vector(-1, 0){500}}
\put(2600,-2686){\vector(-1, 0){500}}
\put(3901,-2311){\vector(-1, 0){500}}
\put(3901,-2686){\vector(-1, 0){500}}
\put(3901,-3061){\vector(-1, 0){500}}
\put(5026,-2686){\vector(-1, 0){500}}
\put(5026,-3061){\vector(-1, 0){500}}
\put(5026,-3436){\vector(-1, 0){500}}
\put(6526,-3436){\vector(-1, 0){500}}
\put(6526,-3811){\vector(-1, 0){500}}
\put(301,-511){\makebox(0,0)[lb]{$[0]$}}
\put(301,-886){\makebox(0,0)[lb]{$[1]$}}
\put(301,-1261){\makebox(0,0)[lb]{$[2]$}}
\put(301,-1636){\makebox(0,0)[lb]{$[3]$}}
\put(301,-2011){\makebox(0,0)[lb]{$[4]$}}
\put(301,-2386){\makebox(0,0)[lb]{$[5]$}}
\put(301,-2761){\makebox(0,0)[lb]{$[6]$}}
\put(301,-3136){\makebox(0,0)[lb]{$[7]$}}
\put(301,-3511){\makebox(0,0)[lb]{$[8]$}}
\put(301,-3886){\makebox(0,0)[lb]{$[9]$}}
\put(226,-4261){\makebox(0,0)[lb]{$\dots$}}
\put(1276,-1261){\makebox(0,0)[lb]{$[2]$}}
\put(1276,-1636){\makebox(0,0)[lb]{$[3]+[1]$}}
\put(1276,-2011){\makebox(0,0)[lb]{$[4]+[2]+[0]$}}
\put(1276,-2386){\makebox(0,0)[lb]{$[5]+[3]+[1]$}}
\put(1276,-2761){\makebox(0,0)[lb]{$[6]+[4]+[2]$}}
\put(1276,-3136){\makebox(0,0)[lb]{$[7]+[5]+[3]$}}
\put(1276,-3511){\makebox(0,0)[lb]{$[8]+[6]+[4]$}}
\put(1276,-3886){\makebox(0,0)[lb]{$[9]+[7]+[5]$}}
\put(1501,-4261){\makebox(0,0)[lb]{$\dots$}}
\put(676,-2386){\makebox(0,0)[lb]{$\dots$}}
\put(2326,-3136){\makebox(0,0)[lb]{$\dots$}}
\put(2776,-1636){\makebox(0,0)[lb]{$[1]$}}
\put(2776,-2011){\makebox(0,0)[lb]{$[2]+[0]$}}
\put(2776,-2386){\makebox(0,0)[lb]{$[3]+[1]$}}
\put(2776,-2761){\makebox(0,0)[lb]{$[4]+[2]$}}
\put(2776,-3136){\makebox(0,0)[lb]{$[5]+[3]$}}
\put(2776,-3511){\makebox(0,0)[lb]{$[6]+[4]$}}
\put(2776,-3886){\makebox(0,0)[lb]{$[7]+[5]$}}
\put(2851,-4261){\makebox(0,0)[lb]{$\dots$}}
\put(3451,-3511){\makebox(0,0)[lb]{$\dots$}}
\put(3901,-3136){\makebox(0,0)[lb]{$[3]+[1]$}}
\put(3901,-2761){\makebox(0,0)[lb]{$[2]+[0]$}}
\put(3901,-2386){\makebox(0,0)[lb]{$[1]$}}
\put(3901,-3511){\makebox(0,0)[lb]{$[4]+[2]$}}
\put(3901,-3886){\makebox(0,0)[lb]{$[5]+[3]$}}
\put(3976,-4261){\makebox(0,0)[lb]{$\dots$}}
\put(4576,-3886){\makebox(0,0)[lb]{$\dots$}}
\put(5026,-2761){\makebox(0,0)[lb]{$[2]$}}
\put(5026,-3136){\makebox(0,0)[lb]{$[3]+[1]$}}
\put(5026,-3511){\makebox(0,0)[lb]{$[4]+[2]+[0]$}}
\put(5026,-3886){\makebox(0,0)[lb]{$[5]+[3]+[1]$}}
\put(5251,-4261){\makebox(0,0)[lb]{$\dots$}}
\put(6526,-3511){\makebox(0,0)[lb]{$[0]$}}
\put(6526,-3886){\makebox(0,0)[lb]{$[1]$}}
\put(6526,-4261){\makebox(0,0)[lb]{$\dots$}}
\put(751,-1186){\makebox(0,0)[lb]{$d_0(2)$}}
\put(751,-1561){\makebox(0,0)[lb]{$d_0(3)$}}
\put(751,-1936){\makebox(0,0)[lb]{$d_0(4)$}}
\put(2350,-1561){\makebox(0,0)[lb]{$d_1(3)$}}
\put(2350,-1936){\makebox(0,0)[lb]{$d_1(4)$}}
\put(2350,-2311){\makebox(0,0)[lb]{$d_1(5)$}}
\put(2350,-2686){\makebox(0,0)[lb]{$d_1(6)$}}
\put(3526,-2311){\makebox(0,0)[lb]{$d_2(5)$}}
\put(3526,-2686){\makebox(0,0)[lb]{$d_2(6)$}}
\put(3526,-3061){\makebox(0,0)[lb]{$d_2(7)$}}
\put(4651,-2686){\makebox(0,0)[lb]{$d_3(6)$}}
\put(4651,-3061){\makebox(0,0)[lb]{$d_3(7)$}}
\put(4651,-3436){\makebox(0,0)[lb]{$d_3(8)$}}
\put(6151,-3436){\makebox(0,0)[lb]{$d_4(8)$}}
\put(6151,-3811){\makebox(0,0)[lb]{$d_4(9)$}}
\end{picture}
}
}
\end{equation}
We leave to the reader a simple task of reconstruction of kernel and images of maps $d_i(j)$.

%
%
%
%
We conclude that the cohomology of the above complex are isomorphic to the algebra $A$.

By proposition \ref{P:ghsfa} the algebra $A$ must have $A_{2,1}=H^{2,1}(L,\mathbb{C})$ as a set of generators. By homogeneity the only possible operation compatible with all structures is multiplication with $5$ arguments, which we denote by $\mu_5$. It maps $W^{\otimes 5}\rightarrow W$ and by duality $W^* \rightarrow W^{*\otimes 5}$. Due to the fact that $A$ is cohomology of the Lie algebra $L$ we must have $\mu_5^*:W^*\rightarrow Free_5(\Pi W^*)\subset W^{*\otimes 5}$ ( $Free$ denote a free Lie algebra). An easy manipulation with $\mathfrak{sl}_2$-representations made with the aid of LiE program shows that $\mu_5^*$ is unique up to a constant. Thus  we conclude that multiplication  must coincide with \ref{E:xdgsy}.

In the table below we tabulated the cohomology of the complex \ref{E:mxbdgdi} for the case of space-time dimension is equal to three.

\begin{equation}
\mbox{
\scriptsize{
$\begin{array}{|c|cccccc}
\dots&\dots&\dots&\dots&\dots&\dots&\dots\\
7&0&0&0&0&0&0\\
6&0&0&0&[0]&0&0\\
5&0&0&0&[1]&0&0\\
4&0&0&0&0&0&0\\
3&0&0&0&0&0&0\\
2&0&0&0&0&0&0\\
1&[1]&0&0&0&0&0\\
0&[0]&0&0&0&0&0\\\hline
&0&1&2&3&4&5
\end{array}$
}
}
\end{equation}
\end{proof}

\begin{definition}
Let $W$ be a two dimensional vector space. In   a graded free Lie algebra $Free(\Pi W)$ generated by odd linear space $\Pi W$ which has grading one define an ideal $I$. The linear space $W$ has a basis $w_{\alpha}$, $\alpha=1,2$. It is also equipped with nondegenerate symplectic bilinear form $\epsilon$, with a matrix $\epsilon_{\beta\gamma}$ in the basis $w_{\alpha}$. 

The ideal  $I$ is generated by relations of the form:
\begin{equation}\label{E:hsiua}
z_{\alpha}=\epsilon^{ \beta \beta'}\epsilon^{ \gamma \gamma'}\{w_{\beta},\{w_{\beta'}\{w_{\gamma},\{w_{\gamma'},w_{\alpha}\}\}\}\}
\end{equation}
\end{definition}
Let $\tilde A_{\alpha\beta}=\{w_{\alpha},w_{\beta}\}$. It is a basis of $Free(\Pi W)_2$. The vectors $\tilde\chi^{\alpha}=\epsilon^{\alpha\delta}\epsilon^{\beta\gamma}[w_{\beta},A_{\gamma\delta}]$ form a basis of $Free(\Pi W)_3$. 

\begin{proposition}
The algebra $YM$ is a subalgebra of $L=Free(\Pi W)/I$. The embedding is defined by the formula:
\begin{equation}\label{E:rqyhd}
\begin{split}
&A_{\alpha\beta}\rightarrow  \tilde A_{\alpha\beta}\\
&\chi_{\alpha}\rightarrow \tilde \chi_{\alpha}
\end{split}
\end{equation}
\end{proposition}
\begin{proof}
The relations \ref{E:hsiua} live in $Free_5(\Pi W)\subset T(W)$ and is $\mathfrak{sl}_2$-invariant subspace isomorphic to $W$. In the course of the proof of \ref{P:ghsfa} we observed that $Free_5(\Pi W)\cong Free_5(\Pi W^*)$ has a unique $\mathfrak{sl}_2$ subrepresentation isomorphic to $W$. In the same proof we show that the space  $H^2(L,\mathbb{C})$of auxiliary degree two is two dimensional and isomorphic to $A_{7,5}\cong W$ . By proposition \ref{P:lswcd} we conclude that $Free(\Pi W)/I$ must coincide with algebra of supersymmetries of $YM$ in dimension three and $YM$ must be it subalgebra. From this we conclude that the map \ref{E:rqyhd} defines an embedding of $YM$.
\end{proof}

\subsection{Relation to the superspace formulation}
Consider a space of coinvariants $\Sym^2(Free(\Pi W))_{Free(\Pi W)}$. We have an element  $\epsilon^{\alpha \beta}z_{\alpha}\circ w_{\beta}$ ($\circ$ stands for symmetric product) which is equal up to a sign to
\begin{equation}\label{E:jctyws}
\epsilon^{\gamma \gamma'}\epsilon^{\alpha \beta}\epsilon^{\alpha' \beta'}\{w_{\alpha},\{w_{\beta},w_{\gamma}\}\}\circ \{w_{\alpha'},\{w_{\beta'},w_{\gamma'}\}\}\in \Sym^2(Free(\Pi W))_{Free(\Pi W)}
\end{equation}

$N=1$ supersymmetric YM theory in dimension 3 admits superspace formulation. Consider a supermanifold $\mathbb{C}^{3|2}$ with coordinates $x_{11},x_{12},x_{22},\theta_1,\theta_2$. The manifold $\mathbb{C}^{3|2}$ is equipped with an odd nonintegrable distribution $G$ of dimension $0|2$. It is spanned by vector fields $\tau_{\alpha}=\pr{\theta_{\alpha}}+\theta_{\beta}\pr{x_{\alpha \beta}}$(with implicit summation over $\beta$). 

It is well know fact in physics(see \cite{DF} for an account for mathematicians), that there is one-to-one correspondence between connections on $\mathbb{C}^{3|2}$ with gauge group $GL_n$ and zero curvature along $G$ and a pair $\nabla, \chi^{\alpha}$ - $GL_n$ connection, spinor field in adjoint representation of $GL_n$.
 
This can be interpreted in terms of algebra $A$. Indeed consider a truncated version  $A^t=\mathbb{C}+W$ of $A$. We have zero multiplication on $W$. The Koszul complex of $A^t$ is $A^t\otimes \Lambda[\theta_1,\theta_2]$ and the differential $d=\lambda^{\alpha}\pr{\theta_{\alpha}}$. The elements $\lambda^{\alpha}$ form a basis of $A_{2,1}$.

On the superspace consider (to simplify even further ) connections which are invariant with respect to $\mathbb{C}^3$ translations. To make everything explicit let us choose a trivialization of the vector bundle $E$. According to \cite{DF} connection is uniquely determined by it restriction to $G$. Denote by ${\cal D}_{\alpha}$ the differentiation of module of sections of $E$ along $\tau_{\alpha}$. In our case translation invariant part of $E$ is a free $\Lambda[\theta_1,\theta_2]$-module of rank $n$. Then ${\cal D}_{\alpha}=\tau_{\alpha}+A_{\alpha}$. Consider an element
\begin{equation}
\lambda^{\alpha}{\cal D}_{\alpha}w^{\alpha}=d+\lambda^{\alpha}A_{\alpha}
\end{equation}
 The condition $(\lambda^{\alpha}{\cal D}_{\alpha})^2=0$, which holds trivially, is equivalent to Maurer-Cartan in such algebra.

\begin{proposition}

There is one-to-one correspondence between classes of gauge equivalent $G$-flat $GL_n$ connections and gauge equivalence classes of solutions of MC equation with values in $Mat_n$.
 A notion of gauge equivalence of $G$-flat connections defined by ${\cal D}_{\alpha}$ on the level of MC-equation transforms into gauge equivalence of solutions of MC equation
\end{proposition}
We omit the proof of this proposition only mention that a good choice of representatives of classes of solutions of MC equation is
\begin{equation}
A_{\alpha\beta}\lambda^{\alpha}\theta^{\beta}+\epsilon_{\alpha\beta}\chi^{\alpha}\lambda^{\beta}\theta^1\theta^2
\end{equation}
We interpret $A_{\alpha\beta}$ as coefficients of connection on $\mathbb{C}^3$ and $\chi^{\alpha}$ as a spinor field.

According to \cite{DF} it is possible to formulate the YM-theory in terms of partial connections ${\cal D}_{\alpha}$. To do that define a quantity $\nabla_{\alpha\beta}=\{{\cal D}_{\alpha},{\cal D}_{\beta}\}$ and
\begin{equation}
\Lambda_{\alpha}=\epsilon^{\gamma \delta}\{{\cal D}_{\gamma}\{{\cal D}_{\delta},{\cal D}_{\alpha}\}\}
\end{equation}
Then the Lagrangian density is equal to
\begin{equation}\label{E:jfgh}
{\cal L}(\Lambda)=\epsilon^{\alpha\alpha'}<\Lambda_{\alpha},\Lambda_{\alpha'}>
\end{equation}
If we expand the formula \ref{E:jfgh} in ${\cal D}_{\alpha}$  using  the definition of $\Lambda_{\alpha}$  we arrive to \ref{E:jctyws} , which coincides with \ref{E:jfgh} with the replacement of $w_{\alpha}$ by ${\cal D}_{\alpha}$.

Let us make this identification more precise. We can formulate superspace version of YM in BV terms. For any field of the theory we should add an antifield. On the level of our algebras it amounts to transition from $A^t\otimes \Lambda[\theta_1,\theta_2]$ to $A\otimes \Lambda[\theta_1,\theta_2]$. Varying  the action \ref{E:jfgh} we make some standard algebraic manipulations with the density $\epsilon^{\alpha \alpha'}<\Lambda_{\alpha},\Lambda_{\alpha'}>$. One of them is integration by parts. In algebraic language it is taken care of by coinvariants in \ref{E:jctyws}. The Euler-Lagrange  equations of the system is precisely $\mu_5=0$ with substitution of the arguments by $w^{\alpha}{\cal D}_{\alpha}$. The such is MC equation for the algebra $A\otimes \Lambda[\theta_1,\theta_2]$ with values in $Mat_n$. From it it is easy to read off the structure maps of the algebra  $A\otimes \Lambda[\theta_1,\theta_2]$, which define such equation. We leave the explicit computation of structure maps of  $A\otimes \Lambda[\theta_1,\theta_2]$ to the interested reader. To summarize above discussion we formulate the following statement whose proof has been just sketched
\begin{proposition}
The space of  solutions of MC-equation for $A\otimes \Lambda[\theta_1,\theta_2]$ with values in $Mat_n$ is tautologically identified with  the space of solutions of  equation of motion  in a mix of superspace and BV formalisms for N=1 D=3 YM theory reduced to a point
\end{proposition}
It is easy to generalize this statement to a full theory, bringing back dependence on space coordinates. This is left to the reader.
\section{Proofs}\label{S:xdqde}
\begin{pf}{\bf \ref{R:kahdb}}
We need to check that the differentiations $\theta_{\alpha}$ transform ideal generated by $\tilde A_i,\tilde \psi_{\alpha}$ into itself. If we prove that $\theta_{\alpha}$ transforms generators $\tilde A_i,\tilde \psi_{\alpha}$ into elements of the  ideal, we would be done. We start with relation $\tilde\psi_{\alpha}$ defined by equation (\ref{E:jhda2})
\begin{align}
&\theta_{\gamma}(\tilde \psi_{\alpha})=\\
&=\Gamma_{\alpha\beta}^i\Gamma_{\gamma\delta i}\{\psi^{\delta},\psi^{\beta}\}+\label{E:sdfwt}\\
&-\frac{1}{2}\Gamma_{\alpha\beta}^i\Gamma_{\gamma}^{\beta kl}[A_i,[A_k,A_l]]\label{E:dwicv}
\end{align}
Due to the symmetry $\{\psi^{\delta},\psi^{\beta}\}=\{\psi^{\beta},\psi^{\delta}\}$ and the  identity \ref{E:afdnc} we can rewrite \ref{E:sdfwt} as
\begin{equation}\notag
\Gamma_{\alpha\beta}^i\Gamma_{\gamma\delta i}\{\psi^{\delta},\psi^{\beta}\}=\frac{1}{2}(\Gamma_{\alpha\beta}^i\Gamma_{\gamma\delta i}+\Gamma_{\alpha\delta}^i\Gamma_{\gamma\beta i})\{\psi^{\delta},\psi^{\beta}\}=-\frac{1}{2}\Gamma_{\alpha\gamma}^i\Gamma_{\delta\beta}^i\{\psi^{\delta},\psi^{\beta}\}
\end{equation}
Using Jacoby identity  we have:
\begin{equation}\label{E:xsjx}
\frac{1}{2}\Gamma_{\alpha\beta}^i\Gamma_{\gamma}^{\beta kl}[A_i,[A_k,A_l]]=\frac{1}{2}(\Gamma_{\alpha\beta}^r\Gamma_{\gamma}^{\beta pq}-\Gamma_{\alpha\beta}^q\Gamma_{\gamma}^{\beta pr})[A_p,[A_r,A_q]]
\end{equation}
Applying identity \ref{E:bsgxv} to the LHS of \ref{E:xsjx} we get an equation for $\Gamma_{\alpha\beta}^i\Gamma_{\gamma}^{\beta kl}[A_i,[A_k,A_l]]$. After solving it we get
\begin{equation}\notag
\frac{1}{2}\Gamma_{\alpha\beta}^i\Gamma_{\gamma}^{\beta kl}[A_i,[A_k,A_l]]=\Gamma_{\alpha\beta}^l[A_i,[A_i,A_l]]
\end{equation}
Thus
\begin{equation}\label{E:safxg}
\theta_{\gamma}(\tilde{\psi}_{\alpha})=\Gamma_{\gamma\alpha}^l\tilde A_l
\end{equation}

Let us understand the action of supersymmetries on the second set of relations $\tilde A_l$.  We split relation $\tilde A_l$ into two pieces for convenience. The first one  contains spinors, the second does not.
\begin{equation}\notag
\theta_{\alpha}([A_i,[A_i,A_k]])=\Gamma_{\alpha\beta i}[[\psi^{\beta},A_i],A_k]+2\Gamma_{\alpha\beta i}[A_i,[\psi^{\beta},A_k]]+\Gamma_{\alpha\beta k}[A_i,[A_i,\psi^{\beta}]]
\end{equation}

\begin{equation}\label{E:waddkf}
\theta_{\alpha}(\frac{1}{2}\Gamma_{\gamma\delta }^k\{\psi^{\gamma},\psi^{\delta}\})=\Gamma_{\gamma\delta }^k\Gamma_{\alpha}^{\delta st}[[\psi^{\gamma},A_s],A_t]
\end{equation}
 Use identity \ref{E:bsgxv} several times it simplify \ref{E:waddkf} and get
\begin{equation}\notag
\Gamma_{\alpha}^{\delta tk}\Gamma_{\delta\gamma}^{s}[[\psi^{\gamma},A_s],A_t]+\Gamma_{\alpha\gamma }^s[[\psi^{\gamma},A_s],A_k]-2\Gamma_{\alpha\gamma }^s[[\psi^{\gamma},A_k],A_s]+\Gamma_{\alpha\gamma }^k[[\psi^{\gamma},A_s],A_s])
\end{equation}
Finally
\begin{equation}\label{E:csdsj}
\theta_{\gamma}(\tilde A_k)=\Gamma_{\gamma}^{\delta tk}[\tilde \psi_{\delta},A_t]
\end{equation}

\end{pf}
\begin{pf}{\bf \ref{P:dxcj}}

The irreducible SUSY data $(S,\Gamma,V)$ has the following property. In the complex vector space $V$ choose a vector $(v,v)\neq0$ and decompose $V=<v>+V'$, with $V'=<v>^{\perp}$. The the map $\Gamma :\Sym^2S\rightarrow <v>+V'$ has two components:$\Gamma(a,b)=<a,b>v+\Gamma'(a,b)$. The symmetric bilinear form  $<.,.>$ is nondegenerate. It could be proved by inspection of irreducible SUSY data in table \ref{T:sdjk}. In the course of the proof one should use the fact that the number of types of  isotypical components in SUSY data does not increase upon restriction of spinor representation from $SO(2n)$ to $SO(2n-1)$. The same number could only double in transition  from $SO(2n+1)$ to  $SO(2n)$. Using the form $<.,.>$ we can low and raise spinor indecies of various tensors. $\Gamma^{'i}_{\alpha\beta}$ is an example of such. Without the form $<.,.>$ even in presence of form $(.,.)$ on $V$ lowering and raising Greek (spinor) indecies is not well defined operation.

Suppose $A_1=v, (v,v)=1$ and $A_2,\dots,A_n$ is an orthogonal basis of $V'$. As we know from \cite{MSch2} the algebra $YM$ is a skew-product
\begin{equation}
YM=<H>\ltimes K(q_1,\dots,q_{dim(V)}|p^1,\dots,p^{dim(V)},\psi^1,\dots,\psi^{dim(S)})
\end{equation}
The algebra $K(q_1,\dots,q_{dim(V)}|p^1,\dots,p^{dim(V)},\psi^1,\dots,\psi^{dim(S)}) $ is by definition a quotient of a free Lie algebra $F(q_1,\dots,q_{dim(V)},p^1,\dots,p^{dim(V)},\psi^1,\dots,\psi^{dim(S)})$, with $q_i,p^j$ even and $\psi^{\alpha}$-odd ($\psi^{\alpha}$ is an orthogonal $<.,.>$-basis of $S$). The ideal is generated  by relation $[q_i,p^i]+\frac{1}{2}\{\psi^{\alpha},\psi_{\alpha}\}$. We used a bilinear form $<.,.>$ to lower indecies of $\psi^{\alpha}$.

The identification of $YM$  with $<H>\ltimes K$ is the following :
\begin{equation}
\begin{split}
&H\rightarrow A_1\\
&q_i\rightarrow A_{i+1}\\
&p_i\rightarrow [A_1,A_{i+1}]\\
\end{split}
\end{equation}
The formulas for the action of $H$ and supersymmetries on $K$ are
\begin{equation}\label{E:jhsgc}
\begin{split}
&Hq_i=p^i\\
&Hp^m=-[q_i,[q_i,q_m]]+\frac{1}{2}\Gamma^{'m}_{\alpha\beta}\{\psi^{\alpha},\psi^{\beta}\}\\
&H\psi^{\alpha}=-\Gamma^{'(i+1)}_{\alpha\beta}[q_i,\psi^{\beta}]\\
&\theta_{\alpha}q_i=\Gamma_{\alpha\beta (i+1)}\psi^{\beta}\\
&\theta_{\alpha}p_i=[\psi^{\alpha},q_i]-\Gamma'_{\alpha\beta i}\Gamma^{'i}_{\beta\delta}[q_i,\psi^{\delta}]\\
&\theta_{\alpha}\psi^{\beta}=-\Gamma^{\beta1i}_{\alpha}p_i-\frac{1}{2}\Gamma^{\beta(i+1)(j+1)}_{\alpha}[q_i,q_j]
\end{split}
\end{equation}
The formulas \ref{E:jhsgc} suggest that $K\subset YM \subset L$ is a chain of ideals in $L$.


It is easy to describe the structure of the quotient $L/K$. Form the identity \ref{P:xsjhfi}  in $L$, we conclude that $\{\theta_{\alpha},\theta_{\beta}\}=-2<\theta_{\alpha},\theta_{\beta}>H$ in  $L/K$. So we can think of $L/K$ as of Clifford Lie algebra. 
 We computed the homology of $K$ in \cite{MSch2}. They are equal to $H_0(K)=\mathbb{C}$, $H_1(K)=V+V^*+S$,$H_2(K)=\mathbb{C}$.  We need also  the action of $L/K$ on cohomology of $K$. We read off the action on cohomology from the action on homology. The action can be recovered on the later from formulas \ref{E:jhsgc}. Denote the image of a generator $x$ of $K$ in $H_1(K)$ by the same symbol $x$.
Such action is trivial on $H_0,H_2$ and on $H_1$is given by the formulas
\begin{equation}
\begin{split}
&Hq_i=p^i\\
&Hp^m=0\\
&H\psi^{\alpha}=0\\
&\theta_{\alpha}q_i=\Gamma'_{\alpha\beta (i+1)}\psi^{\beta}\\
&\theta_{\alpha}p^i=0\\
&\theta_{\alpha}\psi^{\beta}=-\Gamma^{\beta1i}_{\alpha}p_i
\end{split}
\end{equation}
It is obtained from \ref{E:jhsgc} by dropping all the commutators.

Denote $I$ a graded ideal of $L$. Fix a decomposition into direct sum of graded linear spaces $L=I+L/I$.
Then $\Lambda[L^*]=\Lambda[(L/I)^*]\otimes\Lambda[I^*]$. As usual by $^*$ we denote the dualization in the category of  graded vector spaces. Introduce a filtration of $\Lambda[L^*]$ by the formula
\begin{equation}
F^i\Lambda[L^*]=\bigoplus_{j\geq i} \Lambda^j[(L/I)^*]\otimes\Lambda[I^*]
\end{equation}
The algebra $\Lambda[L^*]$ is equipped with a standard Lie algebra differential $d$.
The fact that $I\subset L$ is an ideal guaranties that $d F^i\Lambda[L^*]\subset F^i\Lambda[L^*]$.
Choose some basis $<e_i>$ of algebra $I$
 and a dual basis $<e^i>$ of $I^*$. $<f_i>$,$<f^i>$ are   base of $(L/I)$ and  $(L/I)^*$ respectively. The differential $d$ can be decomposed as
\begin{equation}
\begin{split}
&d=d_{I}+d_a+d_{b}\quad \mbox{{\rm with}}\\
&d_{I}=a_{i,j}^ke^ie^j\frac{\partial}{\partial e^k}\\
&d_a=b_{ij}^ke^if^j\frac{\partial}{\partial e^k}+b_{ij}^{'k}f^if^j\frac{\partial}{\partial f^k}\\
&d_{b}=c_{ij}^kf^if^j\frac{\partial}{\partial e^k}
\end{split}
\end{equation}
The cohomology of $\Lambda[I^*]$ with respect to $d_{I}$ are called Lie algebra cohomology of $I$ with trivial coefficients. They are   denoted by $H(I,\mathbb{C})$, which we usually will abbreviate to $H(I)$.

The operators $d_a,d_{b}$ act trivially on $F^i/F^{i+1}\Lambda[L^*]$. It mean that $d$ reduces to $d_{I}$ on $F^i/F^{i+1}\Lambda[L^*]$ . The cohomology $d$ in $F^i/F^{i+1}\Lambda[L^*]$ are equal to $\Lambda[(L/I)^*]\otimes H(I)$. 

If one applies $d_a+d_{b}$ to any $d_{I}$-cocycle $x\in \Lambda^i[(L/I)^*]\otimes H^j(I)$, lifted to an element of $\Lambda[L^*]$ one gets an element $d_2(x)\in \Lambda^{i+1}[(L/I)^*]\otimes H^j(I)$, which depends only on $d_a$. 

Thus on cohomology of differential $d_1=d_{I}$ there is a structure of a complex with differential $d_2=d_a$. In fact this pattern goes on. A  sequence of complexes $(E_n,d_n)$, with $E_{n+1}=H(E_n)$ is called a spectral sequence. The complex $E_n$ is called the $n$-th term of the spectral sequence. In our case $E_1=\Lambda[(L/I)^*]\otimes H(I)$.

The spectral sequence associated with filtration $F^i$ bears the name of Serre and Hochschild (see \cite{Fuchs} for generalities on Lie algebra cohomology). It is useful because its limiting term $E_{\infty}$ is equal to $H(L)$. We plan to apply the sequence to two ideals $K$ and $YM$ of $L$.

We start with $K$. 

The $E_1$-term of the spectral sequence is equal to $H(K)\otimes \Lambda[H^*]\otimes \Sym(S^*)$
The differential is defined from the action of Clifford Lie algebra on cohomology $H(K)$ and is equal
\begin{equation}
\begin{split}
&dH^*=\theta^{\alpha}\theta^{\alpha}\\
&dp_i^*=q^*_iH^*-\Gamma_{\alpha}^{'\beta1i}\psi^*_{\beta}\theta^{*\alpha}\\
&d\psi^*_{\beta}=\Gamma'_{\alpha\beta i}q^*_i\theta^{*\alpha}
\end{split}
\end{equation}
The two dimensional generator $\lambda\in H^2(K)$ satisfies $p^*_{i}q^{*j}=\delta_i^j\lambda$ $\psi^*_{\alpha}\psi^*_{\beta}=\delta_{\alpha}^{\beta}\lambda$. The action of the differential $d$ on $\lambda,\theta^{*\alpha},q^{*j}$ is trivial.

If we set $\theta^{*\alpha}$ equal to zero then we recover $E_1$ of the spectral sequence from \cite{MSch2} that converges to cohomology of $YM$. In particular the classes that contribute to cohomology of $YM$ are
\begin{align}
&1\mbox{ of cohomological degree one,}\notag \\
&q^*_i,\psi^*_{\beta},H^* \mbox{ of degree one,}\notag\\
&p_i^*H^*,\lambda,\psi^*_{\beta}H^* \mbox{ of degree two and }\label{C:fsdst}\\
&H^*\lambda \mbox{ of degree three. }\notag
\end{align}
We however work in the case when  $\theta^{*\alpha}$ are not zero. It is easy to compute the action of the differential of $E_1$-term  on the classes \ref{C:fsdst}
\begin{align}
&d(p_i^*H^*)=-\Gamma^{\beta1(i+1)}_{\alpha}\psi_{\beta}^*H^*\theta^{*\alpha}-p_i^*\theta^{*\alpha}\theta^{*\alpha}\notag\\
&d(\psi^*_{\beta}H^*)=d(p_i^*H^*\Gamma_{\alpha\beta}^{'i})+\Gamma^{\delta1(i+1)}_{\gamma}\Gamma^{'i}_{\alpha\beta}\psi^*_{\delta}\theta^{*\gamma}\theta^{*\alpha}+\psi^*_{\beta}\theta^{*\alpha}\theta^{*\alpha}\notag\\
&d(\lambda H^*)=\lambda \theta^{*\alpha}\theta^{*\alpha}\label{E:zxzxs}
\end{align}

%
%
%

We turn to a sequence associated with $YM$.
Elements $1,A^*_i,\psi^*_{\alpha},\mu^{*\beta},B^{*j},\omega^*$ are represented by some cocycles in $\Lambda[YM^*]$.

We identify $A^*_i$, $\psi^*_{\alpha}$ with functionals dual to generators $A_i$, $\psi^{\alpha}$. They are $d_{YM}$ cocycles automatically. 

Denote $F^{ij}$ ($F^{ij}=-F^{ji}$) a basis of a linear space dual to space of commutators $<[A_i,A_j]>$.


The formulas for $\mu^{\alpha},B_j,\omega$:
\begin{align}
&\mu^{*\alpha}=\Gamma^{\alpha\beta}_iA^*_i\psi^*_{\beta}\in \Lambda^2[YM^*]\\
&B^*_j=A^{*i}F^{ij}+\Gamma^{\alpha\beta}_j\psi^*_{\alpha}\psi^*_{\beta}\in \Lambda^2[YM^*]\\
&\omega^*=A^*_iA^*_jF^{ij}+\Gamma^{\alpha\beta}_iA^*_i \psi^*_{\alpha}\psi^*_{\beta}\in \Lambda^3[YM^*]
\end{align}

One can consider $\Lambda[YM^*]$ as a subalgebra of $\Lambda[L^*]$. The former algebra is not closed under differential $d$. We use it to relate cohomology of $d$ to cohomology of the complex $A$.

The following formulas will be useful
\begin{align}
&dA^{*i}=\eta_0 A^{*i}\\
&d\psi^{*\alpha}=\eta_1 \psi^{*\alpha}\\
&d\mu^{*\alpha}=\Gamma_i^{\alpha\beta}\Gamma^i_{\gamma\delta}\theta^{*\gamma}\theta^{*\delta}\psi_{\beta}-\Gamma_i^{\alpha\beta}A^{*i}\Gamma_{j\beta\gamma}A^{*j}\theta^{*\gamma}=\\
&\Gamma_i^{\alpha\beta}\Gamma^i_{\gamma\delta}\theta^{*\gamma}\theta^{*\delta}\psi_{\beta}-d(\Gamma_i^{\alpha\beta}\Gamma_{j\beta\gamma}F^{ij}\theta^{*\beta})+\theta^{*\alpha}\theta^{*\beta}\psi_{\beta}=\eta_2\mu^{*\alpha}-d(\Gamma_i^{\alpha\beta}\Gamma_{j\beta\gamma}F^{ij}\theta^{*\beta})\label{E:rfhd}
\end{align}

The action of supersymmetry on $H^2(YM)$ is dual to the action on  $H^1(YM)$ due to Poincare duality. Then the following formula must hold.
\begin{equation}
d(B^{*j\alpha}-\Theta_{\alpha j}\theta^{*\alpha})=\eta_3(B^{*j\alpha})+\tau_{\alpha\beta}\theta^{*\alpha}\theta^{*\beta}
\end{equation}
It is valid for some $\Theta_{\alpha j}$ and $\tau_{\alpha\beta}$.

The ideal $K$ induces a Serre-Hochschild filtration on cohomology $H(YM)$. The leading term of $d\omega^*$ with respect to such filtration is not equal to zero as could be seen from formula \ref{E:zxzxs}.
We conclude
\begin{equation}\label{E:hsgxs}
d(\omega^*-\Theta'_{\alpha }\theta^{*\alpha})=const\Gamma_{\alpha\beta}^i\theta^{*\alpha}\theta^{*\beta}+\tau_{\alpha\beta\gamma}\theta^{*\alpha}\theta^{*\beta}\theta^{*\gamma}
\end{equation}
with some $\Theta'_{\alpha }$, $\tau_{\alpha\beta\gamma}$ and $const\neq 0 $
The module generated $\omega^*$ in $E_1$ term has no torsion over $\Sym(S^*)$. The differential $d_2$ of the spectral sequence nontrivially maps $\omega^*$ into some torsion free module (it is $Ker\eta_3$). Thus $d_2$ is inclusion on such submodule . The higher differential could not act on   $Ker\eta_4\cap\omega^*\Sym(S^*)$ because it is  zero .

Higher differentials of the spectral sequence act trivially on subquotients of $\Sym(S^*)$ modules in $E_1$ generated by $ A^{*i}, \psi^*_{\alpha}$ by dimensional reasons.
Suppose $d_2\mu^{*\alpha}f(\theta)^*_{\alpha}=0$. Then we have
\begin{equation}
d\mu^{*\alpha}f(\theta^*)_{\alpha}=-d\Gamma^{\alpha\beta}_i\Gamma_{\beta\gamma j}F^{ij}\theta^{*\beta}f(\theta^*)_{\alpha}
\end{equation}
Then the element $\mu^{*\alpha}f(\theta^*)_{\alpha}+\Gamma^{\alpha\beta}_i\Gamma_{\beta\gamma j}F^{ij}\theta^{*\beta}f(\theta)_{\alpha}$ is a $d$ cocycle with a leading term with respect to filtration $F^i$ is equal to $\mu^{*\alpha}f(\theta)_{\alpha}$. We conclude that there are no higher differentials on the subquotients of  $\Sym(S^*)<\mu^{*\alpha}>$.

The only possible spot higher differentials could act are subquotients of $\Sym(S^*)<B^{*i}>$. This possibility will be excluded by our explicit computations of cohomology of the complex $A$. It  will be carried out in the following sections. We shall see that the complex $A$ is acyclic in $\Sym(S^*)<B^{*i}>$ term and gives no chance for higher differentials to exist.

From this we conclude that differentials $d_1$ and $d_2$ of the Serre-Hochschild spectral sequence related to a pair $YM\subset L$ could be assembled into complex $A$. Since the spectral sequence collapses in $E_2$ term (no higher differentials) we conclude  that cohomology of $A$ coincide with $H(L)$.
\end{pf}
\begin{remark}
The author believe that if one work hard enough one can avoid introduction of ideal $K$ all the objects related to it. The only reason to introduce it was to prove that the constant $const$ in the formula \ref{E:hsgxs} is not equal to zero. 
\end{remark}
\begin{pf}{\bf \ref{L:ghsgfsd}}
The cohomology ${\cal O}(-i,-i),i>0$ can be computed from the Serre duality.
The canonical class of $\mathbf {P}^3\times \mathbf {P}^1$ is equal to ${\cal O}(-4,-2)$ so $H^4(\mathbf {P}^3\times \mathbf {P}^1,{\cal O}(-i,-i)=H^0(\mathbf {P}^3\times \mathbf {P}^1,{\cal O}(i-4,i-2))^*$.

Fix a linear space $X$. A classical Serres computation tell us that
\begin{equation}\label{Serre}
H^i(\mathbf {P}^n(X^*),{\cal O}(k)=
\begin{cases}
\Sym ^k(X) ,i-0,k\geq 0.\\
\Sym ^{-k+n+1}(X^*),i=n,k\leq -n-1\\
0\mbox{ in all other cases}
\end{cases}
\end{equation}
and Kunneth formula $H^n(Z\times Y,{\cal L}\boxtimes {\cal N})=\bigoplus _{i+j=h}H^i(Z,{\cal L})\otimes H^j(Y,{\cal N})$ (see \cite{GH} for example) shows that $H^0(\mathbf {P}^3\times \mathbf {P}^1,{\cal O}(i,i+2))^*=\Sym ^i (T)\otimes \Sym ^{i+2}W$ and zero for negative value of $i$.
Also $H^l(\mathbf {P}^3\times \mathbf {P}^1,{\cal O}(-k,-k))=0$ for $0<l<4,k>0$.
\end{pf}

\begin{pf}{\bf \ref{P:qwfxjk}}
Suppose we have a map $\mu_{k+l}A^{t\otimes k}\otimes I^{\otimes l}\rightarrow A^t+I$. Pick an element $c_1\otimes  \dots c_k\otimes d_1\otimes \dots \otimes d_l$ with bidegrees of components $(2i_1,i_1),\dots,(2i_k,i_k),(2j_1+6,j_1+4),\dots,(2j_l+6,j_l+4)$. The bidegree of the value of $\mu_{k+l}$ on such element is equal to $(2\sum_{s=1}^ki_s+2\sum^l_{t=1}j_t+6l+2-k-l,\sum_{s=1}^ki_s+\sum^l_{t=1}j_t+4l)$.

For $A^t$-component of such map we must have
\begin{equation}
6l+2-k-l=8l
\end{equation}
For $I$-component  we  have
\begin{equation}
6l+2-k-l+2=8l
\end{equation}

The only possible solution of the first equation is $(k=2,l=0)$. It correspond to multiplication in $A^t$

There are two solutions of the second equation: $(k=1,l=1)$(the module structure) and $(k=4,l=0)$(the ternary map).

The A$_{\infty}$ conditions for $A$ equipped  with  operations $\mu_2,\mu_4$ get translated into  associativity of multiplication in $A^t$, axioms of module on $I$ and cocycle condition on $c$.

The explicit form of the complex \ref{E:mxbdgdi} guaranties that multiplication in $A^t$ and module structure on $I$ are nontrivial. The problem remains to prove nonvanishing of $c$.  If $c$ were equal to zero, then we would have a contradiction with proposition \ref{P:ghsfa}. Indeed there is no way to generate elements of bidegree $(6,4)$ using only multiplication and elements of bidegree $(2,1)$.
\end{pf}

\begin{pf}{\bf \ref{E:hdsvx}}
We contract \ref{E:afdnc} with a tensor $\Gamma^{\alpha\beta}_j\Gamma^{\gamma\delta}_j$. We obtain an equation
\begin{align}
&\Gamma_{\alpha\beta}^i\Gamma_{\gamma\delta}^i\Gamma^{\alpha\beta}_j\Gamma^{\gamma\delta}_j+\label{E:sfdye}\\
&+2\Gamma_{\alpha\gamma}^i\Gamma_{\beta\delta}^i\Gamma^{\alpha\beta}_j\Gamma^{\gamma\delta}_j=0\label{E:lcvsk}
\end{align}
Contracting $\alpha,\beta$ and $i,j$ indecies in \ref{E:javfx} we get that $\Gamma_{\alpha\beta}^i\Gamma^{\alpha\beta}_i=\frac{1}{2}dim(S)dim(V)$. From this and Shur lemma we conclude that $\Gamma_{\alpha\beta}^i\Gamma^{\alpha\beta}_j=\frac{1}{2}dim(S)\delta^i_j$. The last identity can be used to compute \ref{E:sfdye}
\begin{equation}\label{E:zdhesfdye}
\Gamma_{\alpha\beta}^i\Gamma_{\gamma\delta}^i\Gamma^{\alpha\beta}_j\Gamma^{\gamma\delta}_j=\frac{1}{4}dim(S)^2dim(V)
\end{equation}.

We have a following line of identities:
\begin{equation}
\begin{split}
&\Gamma_{\alpha\gamma}^i \Gamma^{\beta\delta}_i \Gamma^{\alpha\beta}_j \Gamma^{\gamma\delta}_j=-\Gamma_{\alpha\gamma}^i \Gamma^{\alpha\beta}_i \Gamma_{\beta\delta}^j \Gamma^{\gamma\delta}_j+ \Gamma_{\alpha\beta}^i \Gamma^{\alpha\beta}_i=\\
&=\Gamma_{\alpha\gamma}^i \Gamma^{\alpha\beta}_i \Gamma_{\beta\delta}^j \Gamma^{\gamma\delta}_j-\delta^{jj}\Gamma_{\alpha\beta}^i \Gamma^{\alpha\beta}_i+\Gamma_{\alpha\beta}^i\Gamma^{\alpha\beta}_i
\end{split}
\end{equation}
Solving this we get
\begin{equation}\label{E:owhsbs}
\Gamma_{\alpha\gamma}^i\Gamma_{\beta\delta}^i\Gamma^{\alpha\beta}_j\Gamma^{\gamma\delta}_j=-\frac{1}{4}(dim(S)dim(V)^2-2dim(S)dim(V))
\end{equation}
Substitution  \ref{E:zdhesfdye} and \ref{E:owhsbs} into \ref{E:sfdye} and \ref{E:lcvsk} we obtain an equation
\begin{equation}
dim(S)dim(V)(\frac{dim(S)}{2}-(dim(V)-2)))=0
\end{equation}
For identity \ref{E:afdnc}  to hold the equation
\begin{equation}\label{E:djsaw}
\frac{dim(S)}{2}=dim(V)-2
\end{equation}
must be satisfied. In physics this equation has the following interpretation. The number $\frac{dim(S)}{2}$ is the number of fermionic degrees of freedom (after equation of motion are taken onto account). The number $dim(V)-2$ are similar bosonic number. Supersymmetry is the statement that these should be equal.

The dimension of a spinor representation in irreducible SUSY data as a function of $dim(V)$ grows exponentially. The formula is given in Appendix. For example if $dim(V)=11$ then $dim(S)=32$ and the equation \ref{E:djsaw} is not satisfied. It is easy to prove inductively (this is left to the reader) that the equation \ref{E:djsaw} has no solution if $dim(V)>10$.

If $dim(V)\leq 10$ it is easy to exhibit by case by case analysis all the solution. They exist for $dim(V)=10,6,4,3$

\end{pf}

\section{Appendix A. Spinor representations and $\Gamma$ matrices}
Suppose $\mathbf{ V}$ is $D$-dimensional vector space, equipped with bilinear symmetric nondegenerate form $(.,.)$. The group $SO(\mathbf{ V})$ with Lie algebra $\mathfrak{ so}(\mathbf{ V})$ acts on $\mathbf{ V}$, preserving $(.,.)$. 

The Lie algebra $\mathfrak{ so}(\mathbf{ V})$ has some special representations which is called spinor representation. The best way to describe them is using Clifford algebra $Cl(\mathbf{ V})$ approach. Denote
\begin{equation}
Cl(\mathbf{ V})=\mathbb{ C}<\mathbf{ V}>/(v_1v_2+v_2v_1-2(v_1,v_2))
\end{equation}
where $v_1,v_2$ runs through the generating space $\mathbf{ V}$. If quadratic form $(.,.)$ is not degenerate then $Cl(\mathbf{ V})$ is a semisimple algebra. This algebra is also $\mathbb{ Z}/2\mathbb{ Z}$ graded with $deg(\mathbf{ V})=1$. One can study $\mathbb{ Z}/2\mathbb{ Z}$-graded representations of such algebra. The Lie group $SO(\mathbf{ V})$ acts on $Cl(\mathbf{ V})$ by automorphisms and define automorphisms of the category of representations. In this category all representations are rigid, this way representations of a double cover of $SO(\mathbf{ V})$, which we denote $Spin(\mathbf{ V})$ appear.

If $D=2n$ then $Cl(\mathbf{ V})$ has an irreducible representation $I$ of dimension $2^n$. This representation is $\mathbb{ Z}/2\mathbb{ Z}$ graded. Then $I=(\mathfrak{ s}_l)_0+(\mathfrak{ s}_r)_1$, where the subscript index $0,1$ denotes parity. The linear space in $I$ of the same parity are invariant with respect to $Spin(\mathbf{ V})$. This way we get two spin representations $\mathfrak{ s}_l,\mathfrak{ s}_r$. The Clifford multiplication defines maps
\begin{equation}
\begin{split}
&\gamma_1:\mathfrak{ s}_l\otimes V\rightarrow \mathfrak{ s}_r\\
&\gamma_2:\mathfrak{ s}_r\otimes V\rightarrow \mathfrak{ s}_l
\end{split}
\end{equation}
Using various parings which exist on spinor representations and which are tabulated in the third column of table (\ref{T:sdjk}) we get maps $\gamma$ which are placed in the forth column of the same table.

It is a general fact that restriction of $\mathfrak{ s}_l, \mathfrak{ s}_r$ from $Spin(2n)$ to $Spin(2n-1)$ remains irreducible. However $\mathfrak{ s}_l|_{Spin(2n-1)}=\mathfrak{ s}_r|_{Spin(2n-1)}=\mathfrak{ s}$. Denote $SO(2n-1)$ equivariant projection $\mathbf{ V}_{2n}\rightarrow \mathbf{ V}_{2n-1}$ by $p$. Then we can get $\gamma$ for $D=2n-1$ by composing any $\gamma_l$ or $\gamma_r$ with $p$

We can think about $\mathbf{ V}$ as of abelian Lie algebra. We would like to describe certain simple odd $\mathfrak{ so}(D)$ equivariant extensions $SUSY$ of $\mathbf{ V}$ such that $\mathbf{ V}\subset$ center $SUSY$. We have an exact sequence of Lie algebras
\begin{equation}
0\rightarrow \mathbf{ V}\rightarrow SUSY \rightarrow \mathbf{ S} \rightarrow 0
\end{equation}
The vector space $\mathbf{ S}$ is purely odd. Anticommutator $\{.,.\}$ in $SUSY$ defines a symmetric map $\Gamma:\Sym^2(\mathbf{ S})\rightarrow \mathbf{ V}$ which will be called $\Gamma$-matrices. The algebra $SUSY$ can be reconstructed from the map $\Gamma$ if we assume that $\Gamma$ is symmetric and $\mathfrak{ so}(D)$ equivariant. Usually one imposes a physical condition that $\mathbf{ S}$ is a sum of spinor representations of $\mathfrak{ so}(D)$. Denote $Ker \Gamma =\{a\in \mathbf{ S}| \Gamma(a,x)=0\ \mbox{{\rm for all }} x\in \mathbf{ S}\}$. It is clear that $Ker \Gamma$ is an ideal in $SUSY$. From now on we assume that $Ker \Gamma=0$.

For a fixed $\mathbf{ V}$ we call a pair $\mathbf{ S},\Gamma$ a SUSY data. SUSY data form a category with obvious definition of a morphism. There is an additive structure on such category:
\begin{equation}
(\mathbf{ S}_1,\Gamma_1)\oplus (\mathbf{ S}_2,\Gamma_2)=(\mathbf{ S}_1+\mathbf{ S}_2,\Gamma_1+\Gamma_2)
\end{equation}
It is easy to describe additive generators in such category.

Let $W$ be a two-dimensional linear space with a symplectic form $\omega$

\begin{equation}\label{T:sdjk}
\mbox{
\setlength{\unitlength}{3947sp}
\begin{picture}(6024,4842)(-11,-4048)
\thinlines
\put(976,764){\line( 0,-1){4800}}
\put(976,-2161){\line( 1, 0){3000}}
\put(976, 89){\line( 1, 0){3000}}
\put(1501,764){\line( 0,-1){4800}}
\put(1501,-4036){\line( 0, 1){ 0}}
\put(2701,764){\line( 0,-1){4800}}
\put(3976,764){\line( 0,-1){4800}}
\put( 1,464){\line( 1, 0){6000}}
\put( 1,-286){\line( 1, 0){6000}}
\put( 1,-661){\line( 1, 0){6000}}
\put( 1,-1411){\line( 1, 0){6000}}
\put( 1,-2536){\line( 1, 0){6000}}
\put( 1,-2911){\line( 1, 0){6000}}
\put( 1,-3661){\line( 1, 0){6000}}
\put( 1,-4036){\line( 1, 0){6000}}
\put(976,-1036){\line( 1, 0){5025}}
\put(976,-3286){\line( 1, 0){5025}}
\put( 1,-1786){\line( 1, 0){6000}}
\put(301,614){\makebox(0,0)[lb]{$dim(\mathbf{ V})$}}
\put(1051,614){\makebox(0,0)[lb]{$\mbox{{\rm spinor}}$}}
\put(1051,484){\makebox(0,0)[lb]{$\mbox{{\rm rep}}$}}
\put(1126,239){\makebox(0,0)[lb]{$\mathfrak{ s}_l$}}
\put(1126,-2461){\makebox(0,0)[lb]{$\mathfrak{ s}_r$}}
\put(376, 89){\makebox(0,0)[lb]{$8n$}}
\put(1126,-2011){\makebox(0,0)[lb]{$\mathfrak{ s}_l$}}
\put(376,-2236){\makebox(0,0)[lb]{$8n+4$}}
\put(1126,-136){\makebox(0,0)[lb]{$\mathfrak{ s}_r$}}
\put(1126,-886){\makebox(0,0)[lb]{$\mathfrak{ s}_l$}}
\put(376,-511){\makebox(0,0)[lb]{$8n+1$}}
\put(1126,-511){\makebox(0,0)[lb]{$\mathfrak{ s}$}}
\put(376,-1111){\makebox(0,0)[lb]{$8n+2$}}
\put(1126,-1261){\makebox(0,0)[lb]{$\mathfrak{ s}_r$}}
\put(376,-1636){\makebox(0,0)[lb]{$8n+3$}}
\put(1126,-1636){\makebox(0,0)[lb]{$\mathfrak{ s}$}}
\put(376,-2761){\makebox(0,0)[lb]{$8n+5$}}
\put(1126,-2761){\makebox(0,0)[lb]{$\mathfrak{ s}$}}
\put(376,-3361){\makebox(0,0)[lb]{$8n+6$}}
\put(1126,-3136){\makebox(0,0)[lb]{$\mathfrak{ s}_l$}}
\put(1126,-3511){\makebox(0,0)[lb]{$\mathfrak{ s}_r$}}
\put(376,-3886){\makebox(0,0)[lb]{$8n+7$}}
\put(1126,-3886){\makebox(0,0)[lb]{$\mathfrak{ s}$}}
\put(1726,614){\makebox(0,0)[lb]{$\mbox{{\rm pairing}}$}}
\put(1651,239){\makebox(0,0)[lb]{$\Sym^2(\mathfrak{ s}_l)\rightarrow \mathbb{ C}$}}
\put(1651,-136){\makebox(0,0)[lb]{$\Sym^2(\mathfrak{ s}_r)\rightarrow \mathbb{ C}$}}
\put(1651,-511){\makebox(0,0)[lb]{$\Sym^2(\mathfrak{ s})\rightarrow \mathbb{ C}$}}
\put(1651,-886){\makebox(0,0)[lb]{$\mathfrak{ s}_l \otimes \mathfrak{ s}_r\rightarrow \mathbb{ C}$}}
\put(1651,-1261){\makebox(0,0)[lb]{$\mathfrak{ s}_r \otimes \mathfrak{ s}_l\rightarrow \mathbb{ C}$}}
\put(1651,-1636){\makebox(0,0)[lb]{$\Lambda^2[\mathfrak{ s}]\rightarrow \mathbb{ C}$}}
\put(1651,-2011){\makebox(0,0)[lb]{$\Lambda^2[\mathfrak{ s}_l]\rightarrow \mathbb{ C}$}}
\put(1651,-2386){\makebox(0,0)[lb]{$\Lambda^2[\mathfrak{ s}_r]\rightarrow \mathbb{ C}$}}
\put(1651,-2761){\makebox(0,0)[lb]{$\Lambda^2[\mathfrak{ s}]\rightarrow \mathbb{ C}$}}
\put(1651,-3136){\makebox(0,0)[lb]{$\mathfrak{ s}_l \otimes \mathfrak{ s}_r\rightarrow \mathbb{ C}$}}
\put(1651,-3511){\makebox(0,0)[lb]{$\mathfrak{ s}_r \otimes \mathfrak{ s}_l\rightarrow \mathbb{ C}$}}
\put(1651,-3886){\makebox(0,0)[lb]{$\Sym^2(\mathfrak{ s})\rightarrow \mathbb{ C}$}}
\put(3001,614){\makebox(0,0)[lb]{$\gamma$}}
\put(2851,239){\makebox(0,0)[lb]{$\mathfrak{ s}_l \otimes \mathfrak{ s}_r\rightarrow \mathbf{ V}$}}
\put(2851,-136){\makebox(0,0)[lb]{$\mathfrak{ s}_r \otimes \mathfrak{ s}_l\rightarrow \mathbf{ V}$}}
\put(2851,-511){\makebox(0,0)[lb]{$\Sym^2(\mathfrak{ s})\rightarrow \mathbf{ V}$}}
\put(2851,-886){\makebox(0,0)[lb]{$\Sym^2(\mathfrak{ s}_l)\rightarrow \mathbf{ V}$}}
\put(2851,-1261){\makebox(0,0)[lb]{$\Sym^2(\mathfrak{ s}_r)\rightarrow \mathbf{ V}$}}
\put(2851,-1636){\makebox(0,0)[lb]{$\Sym^2(\mathfrak{ s})\rightarrow \mathbf{ V}$}}
\put(2851,-2011){\makebox(0,0)[lb]{$\mathfrak{ s}_l \otimes \mathfrak{ s}_r\rightarrow \mathbf{ V}$}}
\put(2851,-2761){\makebox(0,0)[lb]{$\Lambda^2[\mathfrak{ s}]\rightarrow \mathbf{ V}$}}
\put(2851,-3136){\makebox(0,0)[lb]{$\Lambda^2[\mathfrak{ s}_l]\rightarrow \mathbf{ V}$}}
\put(2851,-3511){\makebox(0,0)[lb]{$\Lambda^2[\mathfrak{ s}_r]\rightarrow \mathbf{ V}$}}
\put(2851,-3886){\makebox(0,0)[lb]{$\Lambda^2[\mathfrak{ s}]\rightarrow \mathbf{ V}$}}
\put(4051,614){\makebox(0,0)[lb]{$\mbox{{\rm irreducible SUSY data}}$}}
\put(4126, 89){\makebox(0,0)[lb]{$\mathbf{ S}=\mathfrak{ s}_l+\mathfrak{ s}_r,\Gamma = \gamma_l+\gamma_r$}}
\put(4126,-511){\makebox(0,0)[lb]{$\mathbf{ S}=\mathfrak{ s}, \Gamma = \gamma$}}
\put(4126,-886){\makebox(0,0)[lb]{$\mathbf{ S}=\mathfrak{ s}_l,\Gamma = \gamma_l$}}
\put(4126,-1261){\makebox(0,0)[lb]{$\mathbf{ S}=\mathfrak{ s}_r,\Gamma = \gamma_r$}}
\put(4126,-1636){\makebox(0,0)[lb]{$\mathbf{ S}=\mathfrak{ s}, \Gamma = \gamma$}}
\put(4126,-2236){\makebox(0,0)[lb]{$\mathbf{ S}=\mathfrak{ s}_l+\mathfrak{ s}_r,\Gamma = \gamma_l+\gamma_r$}}
\put(4126,-2761){\makebox(0,0)[lb]{$\mathbf{ S}=\mathfrak{ s} \otimes W,\Gamma = \gamma \otimes \omega$}}
\put(4126,-3136){\makebox(0,0)[lb]{$\mathbf{ S}=\mathfrak{ s}_l \otimes W,\Gamma = \gamma_l \otimes \omega$}}
\put(4126,-3511){\makebox(0,0)[lb]{$\mathbf{ S}=\mathfrak{ s}_r \otimes W,\Gamma = \gamma_r \otimes \omega$}}
\put(4126,-3886){\makebox(0,0)[lb]{$\mathbf{ S}=\mathfrak{ s} \otimes W,\Gamma = \gamma \otimes \omega$}}
\put(2851,-2386){\makebox(0,0)[lb]{$\mathfrak{ s}_r \otimes \mathfrak{ s}_l\rightarrow \mathbf{ V}$}}
\end{picture}
}
\end{equation}
As you can see from the table (\ref{T:sdjk}) for a given $\mathbf{ V}$ the category of $SUSY$ data is generated by either one or two objects.

$dim(\mathbf{ S}_n)$ is the dimension of irreducible SUSY data when dim $dim(V)=n$. The following set of equalities are checked by direct inspection:
\begin{equation}\label{E:hxcds}
\begin{split}
&dim(\mathbf{ S}_{8n+1})=dim(\mathbf{ S}_{8n})\\
&dim(\mathbf{ S}_{8n+2})=dim(\mathbf{ S}_{8n})\\
&dim(\mathbf{ S}_{8n+3})=2dim(\mathbf{ S}_{8n})\\
&dim(\mathbf{ S}_{8n+4})=4dim(\mathbf{ S}_{8n})\\
&dim(\mathbf{ S}_{8n+5})=8dim(\mathbf{ S}_{8n})\\
&dim(\mathbf{ S}_{8n+6})=8dim(\mathbf{ S}_{8n})\\
&dim(\mathbf{ S}_{8n+7})=16dim(\mathbf{ S}_{8n})\\
&dim(\mathbf{ S}_{8n+8})=16dim(\mathbf{ S}_{8n})\\
&dim(\mathbf{ S}_{8n})=16^n
\end{split}
\end{equation}

There is a standard phraseology associated with it. If the category is generated by one object $\mathbf{ S},\Gamma$, then $SUSY$ built by $(\mathbf{ S},\Gamma)^{\oplus k}$ is called $N=k$ supersymmetry group. If the category is generated by two objects $(\mathbf{ S}_1,\Gamma_1)$ and $(\mathbf{ S}_2,\Gamma_2)$ then $SUSY$ built by $(\mathbf{ S},\Gamma)_1^{\oplus k_1}\oplus (\mathbf{ S},\Gamma)_2^{\oplus k_2}$ is called $(N_1,N_2)=(k_1,k_2)$ supersymmetry group.

By inspection it becomes clear that if $\mathbf{ S},\Gamma$ is the only generator of the category then $\mathbf{ S}$ is a selfdual representation. In the case when we have a pair of generators $(\mathbf{ S}_1,\Gamma_1)$ and $(\mathbf{ S}_2,\Gamma_2)$ then $\mathbf{ S}_1$ is dual to $\mathbf{ S}_2$.

In the main body of the paper $\mathbf{ S},\Gamma$ will be some (not necessarily irreducible) $SUSY$ data. Choose a basis $<\psi^{\alpha}>$ of $\mathbf{ S}$ and a basis $<A_i>$ of $\mathbf{ V}$. In this basis the map $\Gamma$ will have a form
\begin{equation}
\Gamma(\psi^{\alpha},\psi^{\beta})=\sum_{i=1}^{D}\Gamma^{\alpha \beta i}A_i
\end{equation}
by the observation of the previous paragraph there is a dual $\Gamma$ map
\begin{equation}
\Gamma(\psi_{\alpha},\psi_{\beta})=\sum_{i=1}^{D}\Gamma_{\alpha \beta i}A_i
\end{equation}
where $<\psi_{\alpha}>$ is a basis of the dual space $\mathbf{ S}^*$.

For any SUSY data $(S,\Gamma,V)$ there is a dual $(S^*,\Gamma,V)$ SUSY data. It could be seen by inspection of table \ref{T:sdjk}. If $(S^*,\Gamma,V)$ is isomorphic to $(S,\Gamma,V)$ as a representations of symmetry group $G(V)$ then we say that $(S,\Gamma,V)$ is selfdual. At any rate the tensors $\Gamma^{i}_{\alpha\beta}$ and $\Gamma_{i}^{\alpha\beta}$ are defined.

The following identity could be considered as defining for $\Gamma$-matrices:
\begin{equation}\label{E:javfx}
\Gamma_{\alpha\gamma}^i\Gamma^{\gamma\beta j}+\Gamma_{\alpha\gamma}^j\Gamma^{\gamma\beta i}=\delta_{\alpha}^{\beta}\delta^{ ij}
\end{equation}
\begin{definition}
Introduce a tensor
\begin{equation}
\Gamma_{\alpha}^{\beta ij}=\Gamma_{\alpha\gamma}^i\Gamma^{\gamma\beta j}-\Gamma_{\alpha\gamma}^j\Gamma^{\gamma\beta i}=2\Gamma_{\alpha\gamma}^i\Gamma^{\gamma\beta j}-\delta_{\alpha}^{\beta}\delta^{ ij}
\end{equation}
\end{definition}
The following is the corollary of the above definition
\begin{equation}\label{E:bsgxv}
\Gamma_{\alpha\beta}^{i}\Gamma_{\gamma}^{\beta kl}+\Gamma_{\alpha}^{\beta ik}\Gamma_{\beta\gamma}^{l}=\Gamma_{\alpha\gamma}^{i}\delta^{kl}-\Gamma_{\alpha\gamma}^{l}\delta^{ik}
\end{equation}
\subsection*{Acknowledgments}
The author would like to thank A. Gorodentsev, A. Khoroshkin, N.Nekrasov, Yu.I. Manin, A.S.Schwarz   for
stimulating discussions. Part of this work was done when the author was staying in  IHES, MPIM. The author  appreciates the
hospitality of these institutions.

\end{document}